\documentclass[11pt,a4paper]{article}
\pdfoutput=1

\usepackage{jcappub}
\usepackage{mathrsfs}
\usepackage{bbm}
\usepackage{bm}
\usepackage{dsfont}
\usepackage{tensor}
\usepackage{xspace}
\usepackage{aeguill}
\usepackage{wasysym}
\usepackage{microtype}
\usepackage{empheq}
\usepackage{ifthen}
\usepackage{amsmath}
\usepackage{subcaption}
\usepackage{wrapfig}

\newcommand{\dd}{{\rm d}}
\newcommand{\pd}[3][]{\frac{\partial^{#1} #2}{\partial {#3}^{#1}}}
\newcommand{\ddf}[3][]{\frac{\dd^{#1} #2}{\dd {#3}^{#1}}}
\newcommand{\ph}{\varphi}

\newcommand{\define}{\equiv}

\newcommand{\ev}[2][]{
\ifthenelse{ \equal{#1}{1} }{ \langle #2 \rangle  }
					{ \ifthenelse{\equal{#1}{2}}{ \big\langle #2 \big\rangle }
						{ \ifthenelse{\equal{#1}{3}}{ \Big\langle #2 \Big\rangle }
							{ \ifthenelse{\equal{#1}{4}}{ \bigg\langle #2 \bigg\rangle }
								{ \ifthenelse{\equal{#1}{5}}{ \Bigg\langle #2 \Bigg\rangle }
									{ \left\langle #2 \right\rangle }
								}
							}
						}
					}
										}
\newcommand{\e}[1]{_\text{#1}}
\newcommand{\h}[1]{^\text{#1}}
\newcommand{\ex}[1]{\mathrm{e}^{#1}}
\newcommand{\U}[1]{\,\text{#1}}

\newcommand{\module}[2][]{
\ifthenelse{ \equal{#1}{1} }{ | #2 |  }
					{ \ifthenelse{\equal{#1}{2}}{ \big| #2 \big| }
						{ \ifthenelse{\equal{#1}{3}}{ \Big| #2 \Big| }
							{ \ifthenelse{\equal{#1}{4}}{ \bigg| #2 \bigg| }
								{ \ifthenelse{\equal{#1}{5}}{ \Bigg| #2 \Bigg| }
									{ \left| #2 \right| }
								}
							}
						}
					}
										}

\newcommand{\delimiters}[4][]{
\ifthenelse{ \equal{#1}{1} }{  #2 #3 #4  }
					{ \ifthenelse{\equal{#1}{2}}{ \big#2 #3 \big#4 }
						{ \ifthenelse{\equal{#1}{3}}{ \Big#2 #3 \Big#4 }
							{ \ifthenelse{\equal{#1}{4}}{ \bigg#2 #3 \bigg#4 }
								{ \ifthenelse{\equal{#1}{5}}{ \Bigg#2 #3 \Bigg#4 }
									{ \left#2 #3 \right#4 }
								}
							}
						}
					}
													}

\newcommand{\p}[2][]{\delimiters[#1]{(}{#2}{)}}
\newcommand{\pac}[2][]{\delimiters[#1]{[}{#2}{]}}
\newcommand{\paac}[2][]{\delimiters[#1]{\{}{#2}{\}}}
\newcommand{\abs}[2][]{\delimiters[#1]{|}{#2}{|}}

\newcommand{\vect}[1]{\boldsymbol{#1}}

\newcommand{\Ups}{\Upsilon}

\newcommand{\di}{\mathrm{d}}

\newcommand{\jacobi}{\mathcal{D}}

\newcommand{\obs}{\text{o}}
\newcommand{\source}{\text{s}}

\newcommand{\cst}{\mathrm{cst}}
\newcommand{\wl}{\mathscr{L}}

\newcommand{\Dd}{\mathrm{D}}

\def\beq{\begin{equation}}
\def\eeq{\end{equation}}
\def\bea{\begin{eqnarray}}
\def\eea{\end{eqnarray}}
\def\bean{\begin{eqnarray*}}
\def\eean{\end{eqnarray*}}
\def \be {\begin{equation}}
\def \ee {\end{equation}}

\title{Geodesic-light-cone coordinates and the Bianchi I spacetime}

\author[a,b]{Pierre Fleury,}
\author[c]{Fabien Nugier,}
\author[d,e]{Giuseppe Fanizza.}

\affiliation[a]{Department of Mathematics and Applied Mathematics, University of Cape Town,\\
Rondebosch 7701, Cape Town, South Africa}
\affiliation[b]{Department of Physics, University of the Western Cape,\\
Robert Sobukwe Road, Bellville 7535, South Africa}
\affiliation[c]{Leung Center for Cosmology and Particle Astrophysics, National Taiwan University,\\
No. 1, Sec. 4, Roosevelt Road, Taipei 10617, Taiwan (R.O.C)}
\affiliation[d]{Dipartimento di Fisica, Universit\`{a} di Bari,\\Via G. Amendola
173, 70126 Bari, Italy}
\affiliation[e]{Istituto Nazionale di Fisica Nucleare, Sezione di Bari,\\Via E. Orabona n. 4, 70125 Bari, Italy}

\emailAdd{pierre.fleury@uct.ac.za}
\emailAdd{fnugier@ntu.edu.tw}
\emailAdd{giuseppe.fanizza@ba.infn.it}

\abstract{The geodesic-light-cone (GLC) coordinates are a useful tool to analyse light propagation and observations in cosmological models. In this article, we propose a detailed, pedagogical, and rigorous introduction to this coordinate system, explore its gauge degrees of freedom, and emphasize its interest when geometric optics is at stake. We then apply the GLC formalism to the homogeneous and anisotropic Bianchi~I cosmology. More than a simple illustration, this application (i) allows us to show that the Weinberg conjecture according to which gravitational lensing does not affect the proper area of constant-redshift surfaces is significantly violated in a globally anisotropic universe; and (ii) offers a glimpse into new ways to constrain cosmic isotropy from the Hubble diagram.}


\date{\today}

\begin{document}

\maketitle

\flushbottom

\section{Introduction}
\label{sec:introduction}

The current standard cosmological model provides an excellent description of the Universe, in agreement with the great majority of observations. This success even allows one to test fundamental physics, from particles to gravitation. However, the outcome of such tests as well as our understanding of the Universe strongly depend on the reliability of the cosmological principle, which states that our Universe is statistically homogeneous and isotropic~\cite{PeterUzan}. It is therefore crucial to independently test those hypotheses.

A sensible approach to this issue consists in describing the cosmological spacetime in terms of coordinates adapted to observations rather than to peculiar symmetries. Examples of such coordinate systems were proposed in the literature since the late 1930s with Temple's \emph{optical coordinates}~\cite{1938RSPSA.168..122T}, followed in the 1970s by Maartens' \emph{observational coordinates}~\cite{Maartens1,Maartens2,1985PhR...124..315E}, and more recently the \emph{geodesic-light-cone (GLC) coordinates}~\cite{P1}. Such a method was used in refs.~\cite{Bester:2013fya,2015MNRAS.453.2364B} to directly reconstruct spacetime's metric from observations. More specifically, the GLC coordinates were first exploited to perform lightcone averages in a perturbed Friedmann-Lema\^itre-Robertson-Walker (FLRW) spacetime, in order to determine the effect of inhomogeneities on the distance-redshift relation~\cite{P2,P4,Marozzi:2014kua}, and therefore on the interpretation of the Hubble diagram~\cite{P3,P5,Ben-Dayan:2014swa}; GLC coordinates were also recently applied to gravitational lensing in general~\cite{P6,Fanizza:2015swa}, to galaxy number counts~\cite{DiDio:2014lka}, and to the propagation of ultra-relativistic particles~\cite{Fanizza:2015gdn}. See also the review~\cite{Nugier:2015awa} by one of us.

In contrast with homogeneity, less effort was dedicated to testing the isotropy of the Universe. Once the so-called bulk flow---i.e. our local motion with respect to the cosmic-microwave-background (CMB) rest frame---is removed, any large-scale anisotropy in the observed sky shall be interpreted as a deviation with respect to the cosmological principle. Such a signal has been tracked within the CMB~\cite{2007MNRAS.380.1387P,2015arXiv150201593P,2016arXiv160401024S,2016arXiv160507178S} or in supernova (SN) data~\cite{2001MNRAS.323..859K,2007A&A...474..717S,2008JCAP...06..018K,2010JCAP...10..018B,2010JCAP...12..012A,2011MNRAS.414..264C,2013A&A...553A..56K,2013PhRvD..87l3522C,2014MNRAS.444.2820S,2015ApJ...810...47J}; so far no significant deviation from global isotropy has been found, but the power of those methods is limited, respectively, by cosmic variance or by the sparsity and inhomogeneity of sky coverage by SN surveys. An alternative approach, based on the cross-correlation between weak-lensing $E$-modes and $B$-modes, has been proposed recently by refs.~\cite{2013PhRvD..87d3003P,2015PhRvD..92b3501P,2015arXiv150301127P}, and should provide an efficient and independent test of cosmic isotropy.

The purpose of the present article is twofold. On the one hand, we propose a pedagogical introduction to the GLC coordinate system, with in particular the first bottom-up derivation of the expression of spacetime's metric with such coordinates. It is also the occasion to explore their geometrical properties and gauge degrees of freedom in details. On the other hand, we use GLC coordinates to determine the optical properties of a simple homogeneous but anisotropic cosmological model, namely the Bianchi~I spacetime. This allows us to analyse the statistical properties of the associated distance-redshift relation, whether it is averaged over the sky or over sources. We will show in particular that the directional average of the squared angular distance is significantly affected by anisotropy.

The article is organised as follows. Section~\ref{sec:GLC} is devoted to the construction of GLC coordinates and to the exploration of their geometry. In sec.~\ref{sec:OpticsGLC}, we review the expression of the main optical quantities---such as the redshift, the Jacobi matrix, and the angular distance---in terms of GLC coordinates. We then turn to their application to the Bianchi~I spacetime in sec.~\ref{sec:Bianchi_GLC}, where we explicitly check the consistency of our results with the literature and discuss the outcome of lightcone averages. We finally conclude and suggest further works in sec.~\ref{sec:conclusion}.

\section{The GLC coordinate system}
\label{sec:GLC}

The geodesic-light-cone (GLC) coordinate system~\cite{P1} consists of four coordinates: a time coordinate~$\tau$ associated with a spacelike foliation of spacetime, a null coordinate~$w$ associated with a null foliation, and two angles $(\theta^a)_{a=1,2}$. Any spacetime geometry can, at least locally, be expressed in terms of those coordinates, whose great interest lies in the fact that they deeply simplify the equation of geometric optics, as we shall see in this section.

\subsection{Geometrical construction}
\label{sec:GLC_construction}

Suppose spacetime geometry allows the existence of an irrotational congruence of timelike geodesics, that we shall call fundamental observers. By virtue of the Fr\"obenius theorem~\cite{2004rtmb.book.....P}, it defines a particular foliation by spacelike hypersurfaces, that we choose to label with a coordinate~$\tau$. In other words, $\tau=\cst$ defines one of those hypersurfaces, and $u_\mu \define -\partial_\mu \tau$ is normal to it. We further impose $u^\mu u_\mu=-1$, so that $u^\mu$ represents the four-velocity of fundamental observers, and $\tau$ their proper time.

Let $\wl_\obs$ be the worldline of an arbitrary fundamental observer, simply referred to as \emph{the observer} in the following. Spacetime can be foliated by the set of all past lightcones of $\wl_\obs$\footnote{A similar construction can be achieved with future lightcones, though less sensible from a cosmological point of view.}. We choose to label them with a coordinate $w$, so that $w=\cst$ defines one such lightcone, i.e. the set of all events from which emitted photons are all received simultaneously by the observer. A natural choice for $w$---though not the only possible one---is thus the reception time~$\tau_\obs$ of a photon emitted from elsewhere towards the observer. Just like for $\tau$, $\partial_\mu w$ defines a normal vector to $w=\cst$ hypersurfaces, which is here proportional to $k_\mu$, the wave four-vector of a photon travelling towards the observer.

The above definitions imply that the intersection of two hypersurfaces $\tau=\cst$ and $w=\cst$ is a spatial two-dimensional surface, which shall be thought of as a sphere whose radius is the travel time of a photon for it to reach $\wl_\obs$. This surface can be parametrised by two angles $(\theta^a)_{a=1,2}$; we define them so that a null geodesic linking an event~$E$ to $\wl_\obs$ is characterised by $\theta^a=\cst$, additionally to the $w=\cst$ condition. The angles $(\theta^a)_{a=1,2}$ are somehow Lagrangian coordinates for photons, a particular choice being the right ascension and declination of the direction in which the observer actually detects them.

\begin{figure}[h!]
\centering
\begin{subfigure}[t]{0.45\linewidth}
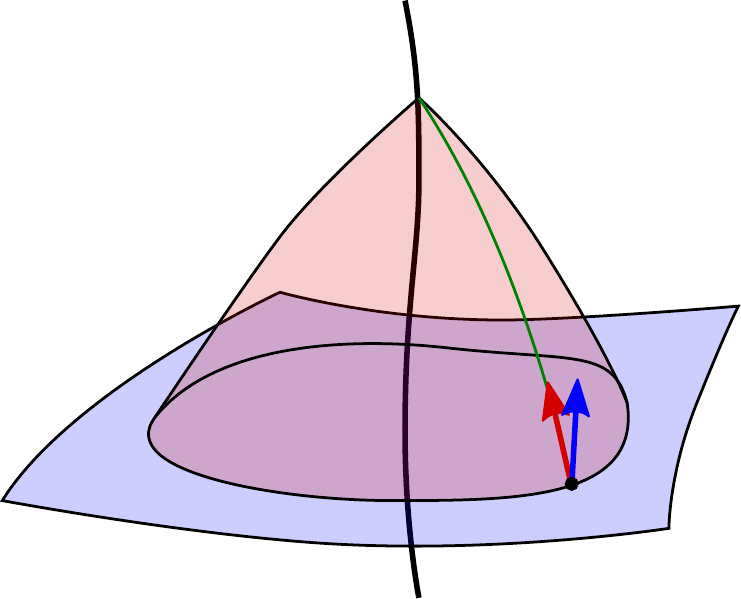
\caption{Definition of the GLC coordinates $\tau,w,\theta^a$. The curve $\wl_\obs$ is the observer's worldline.}
\label{fig:GLC_coordinates}
\end{subfigure}
\hfill
\begin{subfigure}[t]{0.45\linewidth}
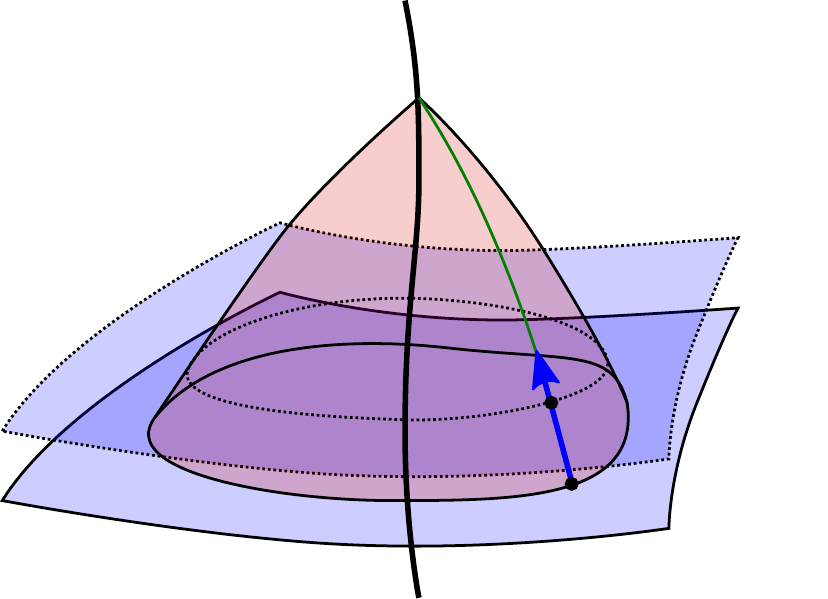
\caption{Basis vector $\vect{\partial}_\tau$ is tangent to the $w,\theta^a=\cst$ lines, hence parallel to $\vect{k}$.}
\label{fig:GLC_dtau}
\end{subfigure}\\[0.5cm]
\begin{subfigure}[t]{0.45\linewidth}
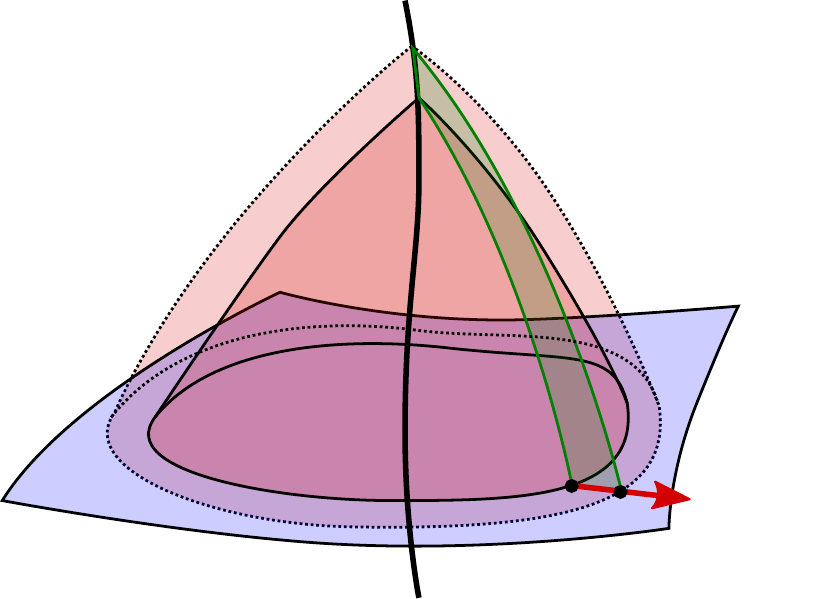
\caption{Basis vector $\vect{\partial}_w$ is tangent to the $\tau,\theta^a=\cst$ lines. It defines a notion of radial direction.}
\label{fig:GLC_dw}
\end{subfigure}
\hfill
\begin{subfigure}[t]{0.45\linewidth}
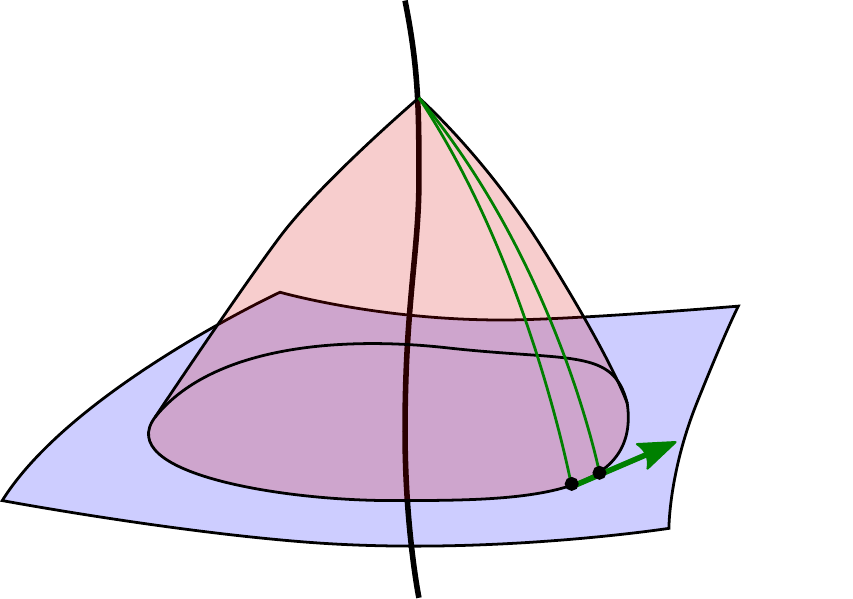
\caption{Basis vectors $\vect{\partial}_a$ are tangent to the $\tau,w,\theta^{b\not=a}=\cst$ lines.}
\label{fig:GLC_dtheta}
\end{subfigure}\\
\caption{Geometry of the GLC coordinates and the associated vector basis.}
\label{fig:GLC}
\end{figure}

All those definitions are summarised in fig.~\ref{fig:GLC_coordinates}. An arbitrary event $E$ is thus coordinated by (i)~the simultaneity hypersurface $\tau$ it belongs to; (ii)~the observer's past lightcone $w$ it is part of; and (iii)~the particular light ray $(\theta^a)$ which connects it to $\wl_\obs$.

Note that the GLC coordinates are unable to describe properly a region of spacetime with caustics, i.e. where a given event~$E$ can be linked with $O\in\wl_\obs$ by several \emph{distinct} light rays, like in strong gravitational lensing systems. Indeed, in such a case $E$ would be described by several sets of angular coordinates~$(\theta^a)$. This would make GLC coordinates ill-defined because not injective. We will not consider this case in the remainder of the article.

\subsection{Coordinate basis and expression of the metric}
\label{subsec:GLC_basis}

Let us now derive the general form of spacetime's metric in GLC coordinates. For that purpose, it is convenient to analyse the properties of the coordinate basis vectors~$\vect{\partial}_\tau,\vect{\partial}_w,\vect{\partial}_a$ associated with $\tau,w,\theta^a$.

By definition, $\vect{\partial}_\tau$ points in the direction of the infinitesimal displacement $E(\tau,w,\theta^a)\rightarrow E'(\tau+\dd\tau,w,\theta^a)$. As illustrated in fig.~\ref{fig:GLC_dtau}, this displacement occurs along the light ray connecting $E$ to $\wl_\obs$, because $(w,\theta^a)$ is unchanged; thus $\vect{\partial}_\tau \propto \vect{k}$, in other words $k^\mu\propto\delta^\mu_\tau$. The proportionality coefficient is determined by noticing that by definition $u_\mu=-\delta^\tau_\mu$, so that
\begin{equation}
k^\tau = -u_\mu k^\mu \define \omega,
\end{equation}
where $\omega$ physically represents the cyclic frequency of the electromagnetic wave associated with $\vect{k}$, as measured by a fundamental observer at $E$. The first basis vector is therefore
\begin{equation}
\vect{\partial}_\tau = \omega^{-1} \vect{k},
\end{equation}
which has two consequences. On the one hand,
\begin{equation}
g_{\tau\tau}
= \vect{g}(\vect{\partial}_\tau,\vect{\partial}_\tau)
= \omega^{-2} \vect{g}(\vect{k},\vect{k})
= 0
\end{equation}
as $\vect{k}$ is a null vector. On the other hand,
\begin{equation}
g_{\tau a}
= \vect{g}(\vect{\partial}_\tau,\vect{\partial}_a)
= \omega^{-1} k_a = 0,
\end{equation}
because $k_\mu \propto \partial_\mu w = \delta^w_\mu$.

The other basis vectors $\vect{\partial}_w$ and $\vect{\partial}_a$ are depicted in figs.~\ref{fig:GLC_dw}, \ref{fig:GLC_dtheta}. The former defines a notion of radial direction, while the latters are tangent to the sphere formed by the intersection between the $\tau=\cst$ hypersurface and the $w=\cst$ lightcone. Note that, by definition, all three are spatial vectors, in the sense that they are orthogonal to $\vect{u}$. At the present stage it is convenient to introduce the usual 3+1 decomposition of the wave four-vector
\begin{equation}
\vect{k} = \omega (\vect{u}+\vect{d}),
\end{equation}
where $\vect{d}$ represents the spatial direction of propagation, in the frame of a fundamental observer, of the photon described by $\vect{k}$. It is a unit vector ($d^\mu d_\mu=1$) orthogonal to $\vect{u}$. It is also orthogonal to $\vect{\partial}_a$, as
\begin{equation}
\vect{g}(\vect{d},\vect{\partial}_a)
=
\vect{g}(\omega^{-1}\vect{k},\vect{\partial}_a)
-
\vect{g}(\vect{u},\vect{\partial}_a)
= 0+0.
\end{equation}

Since the four vectors $\vect{\partial}_w,\vect{\partial}_1,\vect{\partial}_2,\vect{d}$ are all tangent to the $\tau=\cst$ hypersurface, which is three-dimensional, they are not linearly independent, so there exist coefficients $\Upsilon, U^a$ such~that
\begin{equation}\label{eq:Ups_and_U}
\vect{\partial}_w = -\Upsilon\vect{d} - U^a \vect{\partial}_a.
\end{equation}

\begin{figure}[h!]
\centering
\begin{minipage}{6cm}
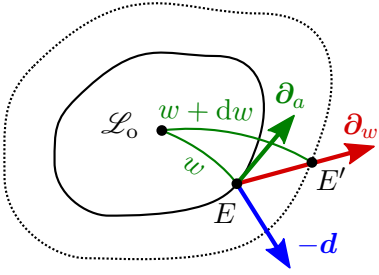
\end{minipage}
\hfil
\begin{minipage}{9cm}
\caption{View ``from above'' of fig.~\ref{fig:GLC_dw}. Two light rays belonging to two successive lightcones $w$ and $w+\dd w$, but with the same angular parameters $\theta^a$ are represented in green. The quantities $U^a$ defined in eq.~\eqref{eq:Ups_and_U} quantify the misalignment between $\vect{\partial}_w$ and the local spatial direction of light propagation~$\vect{d}$. Non-zero $U^a$ are due to both light bending and the rotation of the observer.}
\label{fig:lightcone_shift}
\end{minipage}
\end{figure}

As illustrated in fig.~\ref{fig:lightcone_shift}, $U^a$ quantifies the failure of the parametrisation of light rays on successive lightcones to follow their propagation. We shall thus call it the \emph{photoshift} vector. In terms of $\Upsilon, U^a$, the metric reads
\begin{align}
g_{\tau w}
&= \vect{g}(\vect{\partial}_\tau,\vect{\partial}_w) 
= -\Upsilon,\\
g_{w w}
&= \vect{g}(\vect{\partial}_w,\vect{\partial}_w)
= \Upsilon^2 + g_{ab} U^a U^b,\\
g_{wa}
&= \vect{g}(\vect{\partial}_w,\vect{\partial}_a)
= -g_{ab} U^b \define -U_a,\\
g_{ab} 
&= \vect{g}(\vect{\partial}_a,\vect{\partial}_b)
\define \gamma_{ab},
\end{align}
so that the general expression of the line element in GLC coordinates is finally
\begin{empheq}[box=\fbox]{equation}
\label{eq:GLC_metric}
\dd s^2 = \Ups^2 \di w^2-2\Ups \di w \di \tau+\gamma_{ab}(\di \theta^a-U^a \di w)(\di\theta^b-U^b \di w),
\end{empheq}
in agreement with ref.~\cite{P1}. Under a matrix form, the metric and its inverse in $\tau$, $w$, $\theta^a$ coordinates read
\begin{equation}
\label{GLCmatrix}
[g^{\rm GLC}_{\mu\nu}] =
\begin{bmatrix}
0 & - \Ups &  0 \\
-\Ups & \Ups^2 + U^2 & -(U_b) \\
0  & -(U_a)^{\rm T}  & \gamma_{ab} \\
\end{bmatrix},
\qquad
[g_{\rm GLC}^{\mu\nu}] =
\begin{bmatrix}
-1 & -\Ups^{-1} & -\Ups^{-1}(U^b) \\
-\Ups^{-1} & 0 & 0 \\
-\Ups^{-1}(U^a)^{\rm T} & 0 & \gamma^{ab}
\end{bmatrix},
\end{equation}
where $\gamma^{ab}$ is the inverse of $\gamma_{ab}$, in the sense that $\gamma^{ac}\gamma_{cb}=\delta^a_b$, while a T superscript denotes matrix transposition.

The quantities $\Upsilon$, $U_a$, and $\gamma_{ab}$ encode the six degrees of freedom of the metric, they are a priori functions of the four coordinates $\tau, w, \theta^a$. Note however that the definition of $\gamma_{ab}$ imposes that it vanishes on $\wl_\obs$. In other words the GLC coordinate system is singular on the observer's worldline, just as spherical polar coordinates are singular at their origin.

Let us prove this assertion. Consider two events $E(\tau,w,\theta^a)$ and $E'(\tau,w,\theta^a+\dd\theta^a)$, corresponding to the simultaneous emission of two photons received simultaneously by the observer. Suppose these events are very close to $\wl_\obs$, so that spacetime can be considered flat in this region. Then the proper distance from $E$ (or $E'$) to $\wl_\obs$ is the photons' travel time $\tau_\obs-\tau$, where $\tau_\obs$ is the reception time. So the proper separation between $E$ and $E'$ is
\begin{equation}
\dd \ell^2 \underset{E\rightarrow\wl_\obs}{\sim} (\tau_\obs-\tau)^2 \dd\Omega^2,
\end{equation}
which is also, by definition of the GLC coordinates, $\gamma_{ab} \dd\theta^a \dd\theta^b$, whence $\gamma_{ab}=\mathcal{O}(\tau_\obs-\tau)^2$. 

\subsection{Gauge freedom}

In fact, the construction presented in \S~\ref{sec:GLC_construction} does not fully define the GLC coordinates, because we did not specify (i) how exactly lightcones are labelled; and (ii) how the angles are transferred from one lightcone to another. This freedom in the definition of $w$ and $\theta^a$ allows gauge transformations of the GLC metric, i.e. diffeomorphisms which preserve the \emph{form}~\eqref{eq:GLC_metric} of the metric, but change the \emph{functional expression} of the quantities~$\Upsilon, U_a, \gamma_{ab}$.

\subsubsection{Relabelling lightcones}

Suppose one has found GLC coordinates for a given metric and wish to relabel the lightcones as $w\rightarrow w'(w)$. It is clear that the GLC form of the metric is preserved, and that the GLC functions change according to
\begin{equation}
\Upsilon \rightarrow \Upsilon'=\Upsilon\,\ddf{w}{w'},
\qquad
U_a \rightarrow U_a' = U_a\,\ddf{w}{w'}.
\end{equation}
The first of the above relation tells us in particular that it is always possible, by a redefinition of $w$, to set $\Upsilon$ to $1$ along any curve which crosses at most once each lightcone of $\wl_\obs$.

\begin{figure}[!h]
\centering
\begin{minipage}{4cm}
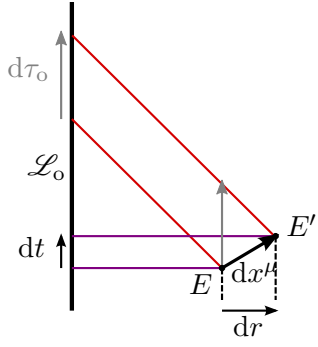
\end{minipage}
\hspace{1cm}
\begin{minipage}{7cm}
\caption{The time interval~$\dd\tau_\obs$ between the receptions of two photons emitted at $E$ and $E'$ is equal to the sum of the time interval~$\dd t$ and the spatial distance~$\dd r$ between those events. This reasoning is only valid in terms of Fermi normal coordinates in the vicinity of $\wl_\obs$.}
\label{fig:dtauo}
\end{minipage}
\end{figure}

Interestingly, imposing this condition on $\wl_\obs$ corresponds to the case $w=\tau_\obs$ where lightcones are labelled by the reception time of their photons. In order to prove this, consider two events $E$ and $E'$ separated by $\dd x^\mu$, assumed to be very close to $\wl_\obs$. Let $t,r$ be such that $(t,r,\theta^a)$ form a Fermi normal coordinate system~\cite{2004rtmb.book.....P} about $\wl_\obs$---note that $t$ coincides with $\tau$ on $\wl_\obs$. The contraction $k_\mu \dd x^\mu$ can then be expressed in two different ways:
\begin{align}
\text{GLC} \qquad k_\mu \dd x^\mu &= k_w^\obs \dd w\label{eq:kdx_GLC}\\
\text{Fermi} \qquad k_\mu \dd x^\mu &= k_t \dd t + k_r \dd r = - \omega_\obs (\dd\tau+\dd r) = -\omega_\obs \dd\tau_\obs,\label{eq:kdx_Fermi}
\end{align}
the last equality being illustrated in fig.~\ref{fig:dtauo}. Besides, the observed frequency is by definition
\begin{equation}
\omega_\obs = -(u^\mu k_\mu)_\obs = - (g^{w\tau} u_\tau k_w)_\obs = - \p{\frac{k_w}{\Upsilon}}_\obs,
\end{equation}
which, once combined with eqs.~\eqref{eq:kdx_GLC}, \eqref{eq:kdx_Fermi}, finally yields
\begin{equation}
\Upsilon|_{\wl_\obs} = \ddf{\tau_\obs}{w}.
\end{equation}
In the following, we shall call \emph{temporal gauge} the choice $w=\tau_\obs$. The articles which have recently introduced and exploited the GLC coordinates \cite{P1,P2,P3,P4,P5,Fanizza:2013doa,Marozzi:2014kua,DiDio:2014lka,P6,Fanizza:2015swa} often give the example of the FLRW geometry, in such a way that $w$ is the conformal time~$\eta_\obs$ of the photon reception by $\wl_\obs$. This choice does \emph{not} correspond to the temporal gauge.

\subsubsection{Relabelling light rays}

Another gauge freedom concerns the way light rays are labelled by the angles $\theta^a$, and how this labelling is transferred from one lightcone to another. Changing the ray-labelling scheme indeed corresponds to a coordinate transformation of the form $\theta^a\rightarrow\ph^\alpha(w,\theta^a)$, which preserves the GLC form of the metric. The quantities~$\gamma_{ab}, U_a$ however change according to
\begin{equation}
\gamma^{ab} \rightarrow \gamma^{\alpha\beta} 
= \gamma^{ab} \partial_a \ph^\alpha \partial_b\ph^\beta,
\qquad
U^a\rightarrow U^\alpha = U^a \partial_a\ph^\alpha - \partial_w \ph^\alpha .
\end{equation}
Since $\ph^\alpha$ is a function of three coordinates $(w,\theta^a)$, the last transformation tells us that it is possible to impose $U^\alpha=0$ on an arbitrary timelike hypersurface, homeomorphic to a cylinder around $\wl_\obs$. This was already noticed in ref.~\cite{Fanizza:2013doa}.

A special case is the limit where this hypersurface is actually $\wl_\obs$, so that $U^\alpha(\wl_\obs)=0$. Geometrically speaking, it means that the labelling of rays locally follows their propagation around $\wl_\obs$. Such a gauge could be baptised \emph{photocomoving}. However, it is important to keep in mind that within such a gauge the angles $\theta^a$ do not necessarily have any observational meaning, namely the directions $\theta^a=\cst$ do not necessarily correspond to fixed directions on the observer's celestial sphere.

An observationally more sensible choice is thus the case where $\theta^a$ are regular spherical angles on the observer's sky at any moment. This choice shall be called the \emph{observational gauge}. If moreover the observer is not rotating, i.e. if any unit vector pointing in a $\theta^a=\cst$-direction is parallely transported along $\wl_\obs$, then we talk about the \emph{non-rotating observational} gauge. In the Minkowski or FLRW spacetimes, the non-rotating observational gauge has $U^a=0$, but as we will see in sec.~\ref{sec:Bianchi_GLC} it is not the case for spacetimes where light rays are not straight lines, such as Bianchi~I.

\subsubsection{Reparametrising light rays}

This paragraph, indirectly related to gauge transformations, deals with the proportionality sign in the relation $k_\mu \propto \partial_\mu w$. In general, the proportionality coefficient~$k_w$ is a function of all coordinates $\tau,w,\theta^a$, but as we will see below it is always possible to reparametrise the null geodesics spanning spacetime so that $k_w$ is a pure constant.

Let $\lambda$ be a parameter on light rays forming the lightcones $w=\cst$, such that $\vect{k}=\dd/\dd\lambda$. For~$\vect{k}$ to physically correspond to a wave four-vector, $\lambda$ cannot be any parameter but an \emph{affine parameter}, which means that $\vect{k}$ satisfies the geodesic equation under its simplest form: $k^\mu \nabla_\mu k_\nu=0$. Independently of the parametrisation, $\vect{k}$ is null so that we can write
\begin{align}
0 &= \frac{1}{2} \nabla_\nu (k^\mu k_\mu) \\
	&= k^\mu \nabla_\nu k_\mu \\
	&= \underbrace{k^\mu (\partial_\nu k_w) \partial_\mu w}_{0} 
			+ k_w k^\mu \underbrace{\nabla_\nu \partial_\mu w}_{\nabla_\mu \partial_\nu w} \\
	&= k^\mu \nabla_\mu (k_w \partial_\nu w) - (k^\mu \partial_\mu k_w) \partial_\nu w \\
	&= k^\mu \nabla_\mu k_\nu - \omega (\partial_\tau k_w) \partial_\nu w.
\end{align}
Therefore, $\lambda$ is an affine parameter iff $\partial_\tau k_w=0$.

In principle, the affine parametrisation can differ from one ray to another. In order to visualise what this means, consider all the rays forming a lightcone~$w=\cst$, suppose their respective affine parameter is fixed to zero on $\wl_\obs$, and consider the set of points corresponding to the (small) value~$\delta\lambda$ of the affine parameter on each of these rays. Those points are therefore separated from $\wl_\obs$ by
\begin{equation}\label{eq:affine_sphere}
\delta x^\mu = k^\mu \delta \lambda = - \frac{k_w \delta\lambda}{\Upsilon_\obs} \, \delta^\mu_\tau,
\end{equation}
hence they form a two-dimensional surface of equation~$w=\cst, \tau=- \Upsilon^{-1}_\obs(w)k_w(w,\theta^a)\delta\lambda$. The affine parametrisation of the different rays can be considered identical if this surface is a sphere, which is the case iff $\partial_a k_w=0$.

A parametrisation of null geodesics such that both the above conditions are fulfilled, i.e. such that $k_w=k_w(w)$, shall be called \emph{isotropic affine parametrisation}. In this case $k_w$ is homogeneous on each lightcones, and by virtue of eqs.~\eqref{eq:kdx_GLC}, \eqref{eq:kdx_Fermi}, equals
\begin{equation}\label{eq:kw_Ups_omegaobs}
k_w = k_w^\obs = -\omega_\obs\,\ddf{\tau_\obs}{w} = -\omega_\obs \Upsilon_\obs.
\end{equation}
Besides, its only non-zero contravariant component is $k^\tau=-\Upsilon^{-1} k_w= \omega_\obs \Ups_\obs / \Ups$. We will see in sec. \ref{sec:OpticsGLC} that this makes $k^\tau$ directly proportional to the redshift $1+z$.

Finally, a similar reasoning can be applied to the way the parametrisation of null geodesics is transferred from one lightcone to another. Suppose we want the events at an affine distance~$\delta\lambda$ from $\wl_\obs$ to be located at the same physical distance from it, whatever the lightcone they belong to. By virtue of eq.~\eqref{eq:affine_sphere}, this imposes that $\partial_w(\Upsilon_\obs^{-1} k_w)=-\partial_w\omega_\obs=0$. This choice shall be called \emph{static affine parametrisation}. 

Summarising, in temporal gauge ($w=\tau_\obs, \Upsilon_\obs=1$) and for a static isotropic affine parametrisation, the only non-zero covariant component $k_w$ of $\vect{k}$ in GLC coordinates is a pure constant. \emph{We choose this gauge for the remainder of this article}.

\subsection{Conformal transformations}
\label{subsec:conformal_transformations}

It is well known that null geodesics are unaffected by conformal transformations of spacetime's metric. We therefore expect the GLC coordinates of a metric to be essentially preserved by conformal transformations. Let $\Omega$ be a scalar function and $\tilde{\vect{g}}$ such that
\begin{equation}
g_{\mu\nu} = \Omega^2 \tilde{g}_{\mu\nu}.
\end{equation}
Suppose one has found a GLC coordinate system $(\tau,w,\theta^a)$ for $\vect{g}$. By definition, the curves $w,\theta^a=\cst$ are null geodesics for $\vect{g}$, and thus for $\tilde{\vect{g}}$ as well. Therefore $w,\theta^a$ are also GLC coordinates for $\tilde{\vect{g}}$. Besides, the coordinate~$\tau$ only needs to be renormalised as $\vect{\partial}_{\tilde{\tau}}=\Omega \vect{\partial}_\tau$ and $\dd\tilde{\tau}=\Omega^{-1}\dd\tau$, so that 
$\tilde{g}_{\tilde{\tau}\tilde{\tau}}
=\tilde{\vect{g}}(\vect{\partial}_{\tilde{\tau}},\vect{\partial}_{\tilde{\tau}})
=\Omega^{-2} \vect{g}(\Omega\vect{\partial}_\tau,\Omega\vect{\partial}_\tau) = g_{\tau\tau}=-1$.

With such definitions, the line element of $\tilde{\vect{g}}$ reads, in terms of~$(\tilde{\tau},w,\theta^a)$,
\begin{align}
\dd\tilde{s}^2 
&= \Omega^{-2} \dd s^2 \\
&= \Omega^{-2}\Upsilon^2 \dd w^2 - 2 \Omega^{-1} \Upsilon \dd w \dd\tilde{\tau}
	+ \Omega^{-2} \gamma_{ab} (\dd\theta^a - U^a \dd w) (\dd\theta^b - U^b \dd w),
\end{align}
from which we immediately deduce that $\Upsilon=\Omega\tilde{\Upsilon}$ and $\gamma_{ab}=\Omega^2 \tilde{\gamma}_{ab}$. Summarising, the GLC coordinates and metric components are affected by conformal transformations according to
\begin{gather}
w=\tilde{w}, \qquad \theta^a=\tilde{\theta}^a, \qquad \dd\tau=\Omega\dd\tilde{\tau}\\
\Upsilon=\Omega\tilde{\Upsilon}, \qquad U^a=\tilde{U}^a, \qquad \gamma_{ab}=\Omega^2 \tilde{\gamma}_{ab}\label{eq:conformal_GLC_metric}.
\end{gather}
However, if one works within the temporal gauge for $\tilde{\vect{g}}$, i.e. with $\tilde{w}=\tilde{\tau}_\obs$, then an additional transformation is required to preserve this gauge with the conformal transformation, namely $\dd w=\dd\tau_\obs=\Omega_\obs \dd\tilde{\tau}_\obs=\Omega_\obs\dd\tilde{w}$. As a consequence, the conformal dictionary becomes
\begin{gather}
\dd w= \Omega_\obs \dd\tilde{w}, 
\qquad 
\theta^a=\tilde{\theta}^a,
\qquad
\dd\tau=\Omega\dd\tilde{\tau}
\label{eq:conformaL_GLC_coord}\\
\Upsilon=({\Omega}/{\Omega_\obs})\tilde{\Upsilon},
\qquad
U^a=\Omega_\obs^{-1}\tilde{U}^a,
\qquad
\gamma_{ab}=\Omega^2 \tilde{\gamma}_{ab}\label{eq:conformal_GLC_metric_temporal_gauge}.
\end{gather}

The main interest of such transformations is that the equations of light propagation are sometimes more easily handled in a conformally-transformed metric~$\tilde{\vect{g}}$, such as in the FLRW or Bianchi~I cases where the scale factor can be absorbed into $\Omega$; the transformation to GLC coordinates is thus more easily performed in this case. The GLC quantities of the original metric~$\vect{g}$ are then recovered from the tilded ones using the dictionary of eqs.~\eqref{eq:conformaL_GLC_coord}, \eqref{eq:conformal_GLC_metric_temporal_gauge}.

\section{Optics with GLC coordinates}
\label{sec:OpticsGLC}

Because they are directly constructed on null geodesics, GLC coordinates are particularly adapted to the analysis of geometric optics in any curved spacetime. This section summarises the GLC expressions of the main optical quantities. In particular, subsec.~\ref{subsubsec:Jacobi_GLC} extends the discussion or ref.~\cite{Fanizza:2013doa} concerning the Jacobi matrix. In this section and the next one, except explicit mention of the contrary, we work in the temporal observational gauge with a static isotropic affine parametrisation.

\subsection{Light rays}

By definition, in GLC coordinates, any light ray is a completely trivial curve with $w,\theta^a=\cst$. Its wave four-vector reads $\vect{k}=\omega \vect{\partial}_\tau$ (see subsec.~\ref{subsec:GLC_basis}) where $\omega$ is the cyclic frequency measured by a fundamental observer. It can also be expressed as
\begin{equation}\label{eq:omega_kw_upsilon}
\omega = -u^\mu k_\mu = -g^{\tau w} u_\tau k_w = -\frac{k_w}{\Upsilon},
\end{equation}
where $k_w$ is a normalisation constant, given our gauge choice. The affine parameter~$\lambda$ along a ray admits a one-to-one relation with the foliation time~$\tau$, namely $\dd\lambda=\omega^{-1}\dd\tau=\Upsilon k_w^{-1} \dd\tau$.

If a light source, following a fundamental worldline ($\vect{u}_\source=\vect{\partial}_\tau$), emits a photon towards the observer, then the redshift~$z$ of this photon reads
\begin{empheq}[box=\fbox]{equation}\label{eq:GLC_redshift}
1+z \define \frac{\omega_\source}{\omega_\obs}= \frac{\Upsilon_\obs}{\Upsilon_\source}.
\end{empheq}
Of course, this expression is only valid for a comoving source, and must be corrected if the latter has a peculiar velocity.\footnote{By `peculiar velocity' we mean here any misalignment between the four-velocitiy of the source with respect to the fundamental flow described by the GLC spatial foliation. Note however that if the flow of sources is smooth and irrotational, then there always exists a foliation such that peculiar velocities are zero.}

\subsection{Light beams}

Let us now turn to infinitesimal light beams, described in terms of the Sachs and Jacobi matrix formalisms, a detailed description of which can be found in, for example, ref.~\cite{Fleury:2015hgz}.

\subsubsection{Sachs basis}
\label{Subsec:Sachs}

The GLC coordinates come with a particular family of observers, with respect to which we choose to define the Sachs basis $(\vect{s}_A)_{A=1,2}$: $u_\mu s_A^\mu=0$ so that $s_A^\tau=0$. Besides, as the Sachs basis spans a screen orthogonal to the line of sight ($k_\mu s_A^\mu=0$) we also have $s_A^w=0$, whence
 \begin{equation}
 \vect{s}_A = s_A^a \vect{\partial}_a.
 \end{equation}

The Sachs basis is an orthonormal basis (zweibein) of the screen space; in other words $\vect{g}(\vect{s}_A,\vect{s}_B)=\gamma_{ab} s_A^a s_B^b=\delta_{AB}$. Introducing the covariant components~$s^A_a\define \vect{g}(\vect{s}_A,\vect{\partial}_a) $, which also form the inverse of $[s_A^a]$ seen as a matrix, the orthonormality relation implies that
\begin{equation}
\gamma_{ab} = \delta_{AB} \, s^A_a s^B_b.
\label{eq:gamma_Sachs}
\end{equation}

For the Sachs basis to be a consistent reference for the orientation of the beam's pattern, it must be parallely transported \emph{as much as possible} along the light beam, while remaining orthogonal to $\vect{u}$ and $\vect{d}$, that is
\begin{equation}
S_{\mu\nu} \frac{\Dd s_A^\nu}{\dd \lambda} = 0,
\end{equation}
where $S^{\mu\nu}\define g^{\mu\nu}-d^\mu d^\nu+u^\mu u^\nu=\delta^{AB}s_A^\mu s_B^\nu$ denotes the screen projector. In terms of GLC components, we have $S_{\mu\nu}=\delta_\mu^a\delta_\nu^b\gamma_{ab}$ so that that the above condition reduces to $\Dd s_A^a/\dd\lambda=0$, or more explicitly
\begin{equation}
\partial_\tau s^a_A = \frac{1}{2} (\gamma^{ac} \partial_\tau \gamma_{bc})s^b_A.
\end{equation}
A general method for solving this differential equation has been proposed in ref.~\cite{Fanizza:2013doa}.

 \subsubsection{Jacobi matrix}
 \label{subsubsec:Jacobi_GLC}
 
It has been shown in ref.~\cite{Fanizza:2013doa} that the Jacobi matrix~$\vect{\jacobi}$, which relates the physical size and shape of a light beam to its observed angular aperture, is directly related to the GLC components of the Sachs basis as
\begin{empheq}[box=\fbox]{equation}
\jacobi\indices{^A_B}(\lambda\leftarrow\lambda_\obs) = s^A_a(\lambda) \, C^a_B,
\label{eq:Jacobi_GLC}
\end{empheq}
where $[C^a_A]$ denotes a $2\times 2$ matrix which is independent from $\lambda$. The initial condition for the derivative of the Jacobi matrix at $\lambda=\lambda_\obs$, namely $(\dd\vect{\jacobi}/\dd\lambda)_\obs=\vect{1}$, implies that the inverse of $[C^a_A]$ reads
\begin{equation}
C^A_a \equiv \p{ \ddf{s^A_a}{\lambda} }_\obs.
\end{equation}
Therefore, once the expression of the Sachs basis in GLC coordinates is found, the Jacobi matrix comes with it for free.

We now extend these results of ref.~\cite{Fanizza:2013doa}, and show that the matrix $[C^a_A]$, as well as its inverse, enjoys a very simple explicit expression. For that purpose, we are going to construct a particular Sachs basis on $\wl_\obs$ from the GLC coordinates themselves. Consider a light ray received by $\wl_\obs$ at $O$, and let $(\vect{e}_\alpha)_{\alpha=0\ldots 3}$, with $\vect{e}_0=\vect{u}$, be a tetrad at this event. We define two angles $(\theta^a)=(\theta,\ph)$ so that the spatial direction of propagation~$\vect{d}_\obs$ of the ray at $O$ reads
\begin{equation}\label{eq:observational_GLC_angles}
-\vect{d}_\obs = \cos\ph \sin\theta\,\vect{e}_1 + \sin\ph \sin\theta\,\vect{e}_2 + \cos\theta\,\vect{e}_3.
\end{equation}
Those angles define observational GLC coordinates, which we can then use to define a particular Sachs basis~$(\bar{\vect{s}}_A)$ at $O$:
\begin{equation}
(\bar{\vect{s}}_1)_\obs = -\partial_\theta \vect{d}_\obs,
\qquad
(\bar{\vect{s}}_2)_\obs = \frac{-1}{\sin\theta} \, \partial_\ph \vect{d}_\obs .
\end{equation}
Such vectors indeed satisfy all the required orthogonality conditions. In the vicinity of $O$, their full GLC expressions are therefore
\begin{equation}
\bar{\vect{s}}_1 = \frac{1}{r}\,\vect{\partial}_\theta + \mathcal{O}(r),
\qquad
\bar{\vect{s}}_2 = \frac{1}{r \sin\theta}\,\vect{\partial}_\ph + \mathcal{O}(r),
\end{equation}
with $r\define\tau_\obs-\tau$. Written in a matrix form, the above relations then provide the GLC components $\bar{s}^a_A$ of the Sachs basis:
\begin{equation}
\begin{bmatrix}
\bar{\vect{s}}_1 & \bar{\vect{s}}_2
\end{bmatrix}
=
\begin{bmatrix}
\vect{\partial}_\theta & \vect{\partial}_\ph
\end{bmatrix}
\cdot
\underbrace{
\frac{1}{r}
\begin{bmatrix}
1 & 0 \\
0 & \frac{1}{\sin\theta}
\end{bmatrix}
}_{[\bar{s}^a_A]}
+ \mathcal{O}(r),
\qquad\text{so}\qquad
[\bar{s}^A_a]=
r
\begin{bmatrix}
1 & 0 \\
0 & \sin\theta
\end{bmatrix}
+
\mathcal{O}(r^3).
\end{equation}
By computing the derivative of the latter matrix at $O$, using $\dd r/\dd\lambda=\omega\dd r/\dd\tau=-\omega$, we finally obtain
\begin{equation}
[\bar{C}^A_a]
= -\omega_\obs
\begin{bmatrix}
1 & 0 \\
0 & \sin\theta
\end{bmatrix},
\qquad
[\bar{C}^a_A]
= -\frac{1}{\omega_\obs}
\begin{bmatrix}
1 & 0 \\
0 & \frac{1}{\sin\theta}
\end{bmatrix}.
\end{equation}
Of course, $(\bar{\vect{s}}_A)$ is not the only possible choice for the Sachs basis: any global rotation with respect to it is admissible too. Thus, the most general expression of matrix~$[C^a_A]$ is
\begin{empheq}[box=\fbox]{equation}
[C^a_A]
=
-\frac{1}{\omega_\obs}
\begin{bmatrix}
\cos\psi & -\sin\psi \\
\sin\psi & \cos\psi
\end{bmatrix}
\begin{bmatrix}
1 & 0 \\
0 & \frac{1}{\sin\theta}
\end{bmatrix},
\label{eq:constant_matrix}
\end{empheq}
where $\psi\in[0,2\pi[$. In principle, the basis could also be reflected, but for simplicity we choose to eliminate this possibility by imposing that the set $(-\vect{d},\vect{s}_1,\vect{s}_2)$ has positive determinant. Besides, as already noticed by ref.~\cite{Fanizza:2013doa}, with generic non-observational GLC angular coordinates~$(\theta^m)$ the above expression must be corrected as $C^m_A=(\partial\theta^m/\partial\theta^a) C^a_A$.

\subsubsection{Angular diameter distance}

The angular diameter distance~$D\e{A}$, which relates the physical size of a source to the angle under which it is seen by the observer, can be deduced from the Jacobi matrix as~\cite{Fleury:2015hgz}
\begin{equation}
D\e{A}^2 = \omega_\obs^2 \, \det\vect{\jacobi}.
\end{equation}
Again, this quantity enjoys a remarkably simple expression in terms of GLC coordinates. Using the formula~\eqref{eq:Jacobi_GLC} for $\vect{\jacobi}$ indeed yields
\begin{equation}
\det\vect{\jacobi} 
= \det[s^A_a] \times \det[C^a_A]
= \sqrt{\det[\gamma_{ab}]} \times \frac{1}{\omega_\obs^2 \sin\theta},
\end{equation}
where in the second equality we used both eqs.~\eqref{eq:gamma_Sachs} and \eqref{eq:constant_matrix}. The GLC expression of the angular diameter distance is therefore
\begin{empheq}[box=\fbox]{equation}
D\e{A}^2 = \frac{\sqrt{\gamma}}{\sin\theta}
\label{eq:GLC_DA}
\end{empheq}
where $\gamma$ denotes the determinant of $[\gamma_{ab}]$. This expression agrees with the recent GLC literature, see e.g. ref.~\cite{NugierThesis} and references therein. Again, a non-observational choice for the angular coordinates can be accounted for by multiplying eq.~\eqref{eq:GLC_DA} with a Jacobian factor~$\det[\partial\theta^m/\partial\theta^a]$.

\subsection{Lightcone averages}
\label{subsec:LCaverage}

The GLC coordinates are particularly adapted to the calculation of the average of cosmological observables~\cite{P1,NugierThesis}. Let $S$ be any scalar observable, such as the luminous intensity or the density of matter. Its \emph{directional average}, i.e. its average over the observer's celestial sphere weighted by solid angles, is naturally defined as
\begin{equation}
\ev{S}_\Omega \define \frac{1}{4\pi} \int S \; \dd^2\Omega,
\end{equation}
with $\dd^2\Omega\define\sin\theta\dd\theta\dd\ph$, and where the angles $\theta,\ph$ are the GLC angles in observational gauge. In the above equation, coordinate~$w$ is fixed as it indicates over which lightcone the average is performed; coordinate~$\tau$ is either fixed to $\tau\e{s}$ if one wishes to average over simultaneous source events, or such that $\tau=\tau(z_\source,\theta,\ph)$ if one is interested in an average over events with the same redshift. Directional averages are typically involved in the analysis of the CMB.

Another important observational notion of average is the \emph{source average}, typically involved in the analysis of cosmological data made involving a catalog of sources, such as the Hubble diagram of SNe, or large galaxy surveys. Contrary to the directional case, it is weighted by the number~$N$ of sources
\begin{equation}
\label{eq:SourceAverage}
\ev{S}_N \define \frac{1}{N} \int S \; \dd^2 N
=
\frac{\int S\; \Sigma \sqrt{\gamma} \, \dd\theta\dd\ph}{\int\Sigma \sqrt{\gamma} \, \dd\theta\dd\ph},
\end{equation}
where we introduced the density of sources per unit of proper area of the source surface, $\Sigma\define \dd^2 N/\dd^2 A = \dd^2 N/(\sqrt{\gamma}\,\dd\theta\dd\ph)$. If this average is performed within a redshift bin, then $\Sigma$ is related to the volumic density of sources~$\rho$ as
\begin{equation}
\Sigma = \rho \, \abs{ \pd{z}{\tau} }^{-1} \,
= \rho \Upsilon^2 \p{\pd{\Upsilon}{\tau}}^{-1}
\end{equation}
in temporal gauge ($\Upsilon_\obs=1$). Note that, introducing the expression~\eqref{eq:GLC_DA} of the angular diameter distance in the definition~\eqref{eq:SourceAverage} of the source average, we get the following identity
\begin{equation}
\ev{ S }_N =
\frac{\int S\; \Sigma D\e{A}^2 \dd^2\Omega}{\int\Sigma D\e{A}^2 \dd^2\Omega}
= \frac{\ev{S\,\Sigma D\e{A}^2}_\Omega}
{\ev{\Sigma D\e{A}^2}_\Omega}.
\end{equation}
%

\section{Bianchi I and GLC}
\label{sec:Bianchi_GLC}

In the literature, the GLC coordinate system has already been applied to the second-order perturbed FLRW spacetime, either in Newtonian gauge~\cite{P2,Fanizza:2015swa,Fanizza:2015gdn}, Poisson gauge~\cite{P5,Marozzi:2014kua,NugierThesis}, or synchronous gauge~\cite{NugierThesis,Fanizza:2013doa}; but also to the Lema\^itre-Tolman-Bondi spacetime~\cite{P6}. This last section is devoted to another non-perturbative application of the GLC coordinates, namely the Bianchi~I geometry.

\subsection{The Bianchi I spacetime}

The Bianchi geometries~\cite{1969CMaPh..12..108E} describe homogeneous but anisotropic cosmological models. Following the Bianchi classification of homogeneous three-dimensional spaces~\cite{LuigiBianchi}, such spacetimes exist under nine different forms, depending on their spatial curvature. The Bianchi~I spacetime, with zero spatial curvature, is the simplest one and also the easiest to interpret: it can be seen as the anisotropic extension of a spatially flat FLRW model, with three different scale factors instead of one. In other words, it describes a homogeneous universe whose expansion is faster in some directions and slower in others. Its metric is conveniently parametrised as
\begin{equation}\label{eq:Bianchi_I_metric}
 \dd s^2 = -\dd t^2 + a^2(t) \gamma_{ij}(t) \dd x^i \dd x^j,
 \qquad
 \gamma_{ij}(t) \define \ex{2\beta_i(t)} \delta_{ij},
\end{equation}
with no summation over $i$ in the definition of $\gamma_{ij}$, and where the three $\beta_i$ must sum to zero,
\begin{equation}\label{eq:constraint}
 \sum_{i=1}^3 \beta_i=0.
\end{equation}
In eq.~\eqref{eq:Bianchi_I_metric}, $t$ denotes cosmic time, which physically represents the proper time of fundamental (or comoving) observers following $x^i=\cst$ worldlines. Those curves are easily shown to be timelike geodesics. The quantity~$a(t)=(\det[g_{ij}])^{1/6}$ is the \emph{volumic scale factor}, which describes the expansion (or contraction) of a 3d domain with constant spatial coordinates. The $\beta_i$ exponents then quantify the volume-preserving deformations of such a domain.

We can also introduce a conformal time~$\eta$ such that $\dd t=a\dd\eta$, so that the volumic scale factor factorises in the expression of the metric, allowing us to easily perform conformal transformations. This will turn out to be convenient in the next subsections.

The anisotropic component of the expansion rate is encoded in the shear rate tensor
\begin{equation}
\sigma_{ij} \define \frac{1}{2} (\gamma_{ij})',
\end{equation}
where a prime denotes a derivative with respect to $\eta$. The indices $i$, $j$ are here conventionally raised by the inverse $\gamma^{ij}$ of $\gamma_{ij}$, so that $\sigma^i_j = \beta'_i \delta^i_j$. The trace of its square is usually denoted by
\begin{equation}
\sigma^2 \define \sigma^{ij} \sigma_{ij} = \sum_{i=1}^3 (\beta'_i)^2.
\end{equation}
For a Bianchi I Universe filled by matter with no anisotropic stress, $\sigma^i_j$ decays as $a^{-2}$, while the analogue of the Friedmann equation governing the evolution of $a$ is
\begin{equation}
\mathcal{H}^2 = \frac{8\pi G\rho}{3} \, a^2 + \frac{\sigma^2}{6}
= \mathcal{H}_0^2\pac{\Omega\e{m0} \p{\frac{a_0}{a}} + \Omega_{\sigma 0} \p{ \frac{a_0}{a}}^4},
\end{equation}
where $\mathcal{H}\define a'/a$ is the conformal expansion rate, $G$ Newton's constant, $\rho$ the density of matter, and a $0$ subscript indicates that the quantity is evaluated today. The matter cosmological parameter~$\Omega\e{m0}$ has the same definition as in standard cosmology, while $\Omega_{\sigma 0} \define \sigma_0^2/(6\mathcal{H}_0^2)$ quantifies the departure from isotropy, we shall therefore call it the \emph{anisotropy parameter}.

Note finally that, as any homogeneous geometry, Bianchi I admits three Killing vectors~$(\vect{\xi}_i)$, which turn out to coincide with the comoving coordinate basis: $\vect{\xi}_i=\vect{\partial}_i$. An interesting consequence is that the scalar product between $\vect{\partial}_i$ and the tangent vector~$\vect{t}$ of any geodesic is then a constant~\cite{Fleury:2015hgz}. This applies in particular to the wave four-vector~$\vect{k}$ of any light ray, so that
\begin{equation}
k_i = \vect{g}(\vect{k},\vect{\partial}_i) = \cst.
\end{equation}
%

\subsection{Coordinate transformation}

Let us now perform the coordinate transformation in order to derive the GLC quantities for the Bianchi~I geometry.

\subsubsection{The conformal trick}

As suggested above, the volumic scale factor~$a(t)$ can be factorised in the expression of the metric by introducing conformal time~$\eta$. This strongly encourages us to perform a conformal transformation of the metric $\tilde{g}_{\mu\nu}\define a^{-2} g_{\mu\nu}$, whose line element is
\begin{equation}
\dd\tilde{s}^2 = -\dd\eta^2 + \sum_{i=1}^3 \pac{\ex{\beta_i(\eta)}\,\dd x^i}^2,
\end{equation}
and therefore get free from $a(t)$. In the remainder of this subsection, we will perform all intermediate calculations working with $\tilde{\vect{g}}$. In other words, we proceed as if we had $a=1$. The full results including the scale factor will however be recovered from the conformal dictionary established in subsec.~\ref{subsec:conformal_transformations}, with $\Omega=a$.

\subsubsection{Choosing the Bianchi GLC coordinates}

We choose the GLC spacelike foliation to coincide with the natural foliation of the Bianchi~I spacetime, so that the GLC time coordinate is $\tilde{\tau}=\eta$. Besides, we work in temporal gauge, hence $\tilde{w}=\eta_\obs$. We did not specify explicitly which worldline is $\wl_\obs$ because Bianchi~I is homogeneous, $\wl_\obs$ can thus be any fundamental observer, e.g. $x^i=0$.

Regarding angular coordinates~$\theta^a$, we work in the observational gauge, and define them from a tetrad~$(\tilde{\vect{e}}_\alpha)_{\alpha=0\ldots 3}$ following $\wl_\obs$ as in \S~\ref{subsubsec:Jacobi_GLC}. Such a tetrad can be
\begin{equation}
\tilde{\vect{e}}_0=\vect{\partial}_\eta,
\qquad
\tilde{\vect{e}}_i=\ex{-\beta_i}\vect{\partial}_i \quad \text{(no summation).}
\end{equation}
It corresponds to a non-spinning observer. The angles $(\theta^a)=(\theta,\ph)$ are then defined from the observed spatial direction of propagation~$\tilde{\vect{d}}_\obs$ of an incoming photon as in eq.~\eqref{eq:observational_GLC_angles}. We call $D^i(\theta,\ph)$ the components of $\tilde{\vect{d}}_\obs$ over the spatial part of the tetrad, i.e. such that
\begin{equation}
\tilde{\vect{d}}_\obs
= D^i(\theta,\ph)\,\tilde{\vect{e}}_i 
= -\sin\theta\cos\ph\,\tilde{\vect{e}}_1 - \sin\theta\sin\ph\,\tilde{\vect{e}}_2 - \cos\theta\,\tilde{\vect{e}}_3.
\end{equation}

\subsubsection{Mapping between comoving and GLC coordinates}

Let us now establish the explicit mapping
\begin{equation}
(\eta,x,y,z) \mapsto (\eta,\eta_\obs,\theta,\ph)
\end{equation}
between comoving and (conformal) GLC coordinates. This exact step has actually been done in the late 60s by Saunders~\cite{saunders_observations_1968,saunders_observations_1969}, in order to derive the distance-redshift relation in the Bianchi~I spacetime. The easiest way to perform the mapping consists in integrating the wave four-vector with respect to the affine parameter:
\begin{equation}\label{eq:x^i}
x^i = \int_{\tilde{\lambda}_\obs}^{\tilde{\lambda}} \dd\tilde{\lambda}' \; \tilde{k}^i
= \int_{\eta_\obs}^\eta \dd\eta' \; \tilde{\omega}^{-1} \tilde{g}^{ij}(\eta) \tilde{k}_j,
\end{equation}
where we used that $\dd\tilde{\lambda}=\tilde{\omega}^{-1}\dd\tilde{\tau}=\tilde{\omega}^{-1}\dd\eta$. We then exploit the fact that $\tilde{k}_i$ is a constant along the light ray, hence it is equal to its observed value
\begin{equation}\label{eq:ki_tilde}
\tilde{k}_i 
= \tilde{k}_i^\obs
= \tilde{\omega}_\obs \tilde{d}^\obs_i
= \tilde{\omega}_\obs \, \tilde{\vect{g}}(\tilde{\vect{d}}_\obs,\vect{\partial}_i)
= \tilde{\omega}_\obs \ex{\beta_i(\eta_\obs)} D^i(\theta,\ph)
\qquad \text{(no summation).}
\end{equation}

Besides, we can use the fact that $\tilde{\vect{k}}$ is null to express the frequency~$\tilde{\omega}$ in terms of $\tilde{k}_i$,
\begin{equation}\label{eq:omega_tilde}
\tilde{\omega}^2 = \tilde{g}^{ij} \tilde{k}_i \tilde{k}_j = \sum_{i=1}^3 \pac{ \ex{-\beta_i(\eta)} \tilde{k}_i }^2.
\end{equation}
Gathering eqs.~\eqref{eq:x^i}, \eqref{eq:ki_tilde}, we finally get the explicit relation between comoving coordinates and (conformal) GLC coordinates,
\begin{empheq}[box=]{equation}
x^i(\eta,\eta_\obs,\theta,\ph) 
= \ex{\beta_i(\eta_\obs)} D^i(\theta,\ph)
	\int_{\eta_\obs}^{\eta} \dd \eta' \; \ex{-2\beta_i(\eta')}
	\paac{ \sum_{j=1}^3 \pac{ \ex{\beta_j(\eta_\obs)-\beta_j(\eta')} D^j(\theta,\ph) }^2 }^{-1/2}.
\label{eq:mapping_observational}
\end{empheq}

In order to derive the GLC form of the Bianchi~I metric, we will need to express $\dd x^i$ as a function of $\dd\eta, \dd\eta_\obs,\dd\theta^a$. For that purpose, let us introduce some useful quantities and relations. The key partial derivatives can be formally expressed as
\begin{align}
\pd{x^i}{\eta} &= \tilde{\omega}^{-1} \tilde{k}^i = \tilde{d}^i, \\
\pd{x^i}{\eta_\obs} &= -\tilde{d}^i_\obs + \int_{\eta_\obs}^\eta \dd\eta'\;\pd{\tilde{d}^i}{\eta_\obs} \define -\tilde{d}^i_\obs + I^i_{\eta_\obs},\\
\pd{x^i}{\theta^a} &= \int_{\eta_\obs}^{\eta} \dd\eta'\;\pd{\tilde{d}^i}{\theta^a} \define I^i_a,
\end{align}
where we have defined three integrals $I^i_{\eta_\obs}, I^i_a$. It is not necessary to calculate them explicitly, but it is worth remarking that they are all orthogonal to $\vect{k}$, in the sense that
\begin{equation}
k_i I^i_{\eta_\obs} = k_i I^i_a = 0.
\label{eq:orthogonality_integrals}
\end{equation}
Indeed, as $k_i=\tilde{k}_i$ is a constant along light rays it can safely enter into the integrals; we are thus interested by terms of the form $k_i \partial \tilde{d}^i$, where $\partial$ denotes a derivative with respect to either $\eta_\obs$ or $\theta^a$. Since $\tilde{g}_{ij}$ only depends on $\eta$ we can write
\begin{equation}
0=\partial \p{\tilde{d}_i \tilde{d}^i} 
= \tilde{g}_{ij}\partial \p{\tilde{d}^i \tilde{d}^j}
= 2 \tilde{d}_i \partial \tilde{d}^i
= 2 \tilde{\omega}^{-1} k_i \partial \tilde{d}^i,
\end{equation}
so that the integrands of $k_i I^i_{\eta_\obs}$ and $k_i I^i_a$ all vanish identically.

\subsubsection{GLC form of the Bianchi I metric}

We now introduce the expression of $\dd x^i$ as a function of GLC coordinates in the line element,
\begin{align}
\dd \tilde{s}^2 &= \tilde{g}_{\mu\nu} \dd x^\mu \dd x^\nu \\
			&= -\dd\eta^2 + \tilde{g}_{ij} \pac[3]{ \tilde{d}^i \dd\eta 
									+ (I^i_{\eta_\obs}-\tilde{d}^i_\obs) \dd\eta_\obs 
									+ I^i_a \dd\theta^a} 
							\pac[3]{ \tilde{d}^j \dd\eta 
									+ (I^j_{\eta_\obs}-\tilde{d}^j_\obs) \dd\eta_\obs 
									+ I^j_b \dd\theta^b} 	\\
&= \tilde{g}_{ij} (I^i_{\eta_\obs}-\tilde{d}^i_\obs) (I^j_{\eta_\obs}-\tilde{d}^j_\obs)  \dd\eta_\obs^2
	- 2 \tilde{d}_i \tilde{k}^i_\obs \, \dd\eta \dd\eta_\obs - 2 \tilde{g}_{ij} (\tilde{d}^i_\obs - I^i_{\eta_\obs}) I^j_a \, \dd\eta_\obs \dd\theta^a + \tilde{g}_{ij} I^i_a I^j_b \, \dd\theta^a \dd\theta^b,
\end{align}
which, by identification with the general form~\eqref{eq:GLC_metric} of the metric in GLC coordinates, yields
\begin{align}
\tilde{\Upsilon} &= \tilde{d}_i \tilde{d}^i_\obs = \tilde{\omega}_\obs/\tilde{\omega},\\
\tilde{U}^2 &= \tilde{g}_{ij} (I^i_{\eta_\obs}-\tilde{d}^i_\obs) (I^j_{\eta_\obs}-\tilde{d}^j_\obs) - (\tilde{\omega}_\obs/\tilde{\omega})^2, \\
\tilde{U}_a &= \tilde{g}_{ij} (\tilde{d}^i_\obs-I^i_{\eta_\obs}) I^i_a,\\
\tilde{\gamma}_{ab} &= \tilde{g}_{ij} I^i_a I^j_b.
\end{align}
Reintroducing the scale factor using the conformal dictionary of subsec.~\ref{subsec:conformal_transformations}, we finally get the following semi-explicit expressions for the GLC quantities of the Bianchi~I spacetime:
\begin{empheq}[box=\fbox]{align}
\Upsilon &=
\frac{a(t)}{a(t_\obs)}
\paac{ \sum_{i=1}^3 \pac{\ex{\beta_i(t_\obs)-\beta_i(t)} D^i(\theta,\ph)}^2}^{-1/2}, \label{eq:Upsilon_Bianchi_I}\\
U_a &= g_{ij} \p{d^i_\obs-I^i_{t_\obs}} I^j_a, \label{eq:U_Bianchi_I}\\
\gamma_{ab} &= g_{ij} I^i_a I^j_b,
\end{empheq}
with $I^i_{t_\obs}\define I^i_{\eta_\obs}/a(t_\obs)$. In order to obtain eq.~\eqref{eq:Upsilon_Bianchi_I}, we introduced the expression~\eqref{eq:ki_tilde} of $\tilde{k}_i$ in the formula~\eqref{eq:omega_tilde} for $\tilde{\omega}$. Besides, in eq.~\eqref{eq:U_Bianchi_I} we used that $d^i_\obs=\tilde{d}^i_\obs/a_\obs$. Following eq.~\eqref{eq:GLC_redshift}, we conclude that the redshift in a Bianchi~I universe reads
\begin{equation}
1+z = \frac{a(t_\obs)}{a(t_\source)} \sqrt{ \sum_{i=1}^3 \pac{\ex{\beta_i(t_\obs)-\beta_i(t_\source)} D^i(\theta,\ph)}^2 },
\end{equation}
in agreement with ref.~\cite{FPU}.

Introducing the inverse of $[I^i_a]$, denoted $[I^a_i]$ and defined by the relations
\begin{equation}
I^i_a I_j^b = \delta^b_a,
\qquad
I^i_a I^a_j = S^i_j \define \delta^i_j - d^i d_j,
\end{equation}
it is easy to show that the photoshift vector reads $U^a=I^a_i (d^i_\obs-I^i_{t_\obs})$. It is not possible to get a more explicit expression. However, a long but straightforward calculation, based on expansions of the various involved quantities around $\wl_\obs$, yields
\begin{equation}
U^a_\obs = - \p{ \dot{g}_{ij} d^i\partial_a d^j }_\obs,
\end{equation}
where a dot denotes a derivative with respect to $t$. Note that this quantity vanishes if $\vect{d}_\obs$ is aligned with a proper axis of $\vect{g}$. This is in agreement with the discussion illustrated in fig.~\ref{fig:lightcone_shift}. Indeed, we have seen that a non-zero value of $U^a$ is due to both light deflection and the rotation of the observer. Here the observer does not spin, but light rays are generally bent, except when they propagate along a proper axis of the expansion.

\subsection{Jacobi matrix}

Let us now check the consistency between the expression of the Jacobi matrix obtained from the GLC formalism~\cite{Fanizza:2013doa} and recent results of the literature \cite{FPU}. In this subsection, we keep working in the conformal geometry~$\tilde{\vect{g}}$. \emph{We however drop all the tildes in order to alleviate notation}. In ref.~\cite{FPU}, the Jacobi matrix in a conformal Bianchi~I spacetime is found to be
\begin{equation}\label{eq:Jacobi_FPU}
\jacobi\indices{^A_B} = s^A_i \mathcal{I}^{ij} (s_{Bj})_\obs,
\end{equation}
where the integral matrix~$\mathcal{I}^{ij}$ is defined by
\begin{equation}
\mathcal{I}^{ij} \define \int_{\eta_\obs}^\eta \dd\eta' \; \omega^{-1} S^{ij},
\end{equation}
with $S^{ij}=g^{ij}-d^i d^j=\delta^{AB} s^i_A s^j_B$ the spatial components of the screen projector introduced in subsec. \ref{Subsec:Sachs}.

In order to connect this result with the GLC quantities that we have just obtained, a first step consists in remarking that, as $\partial_a d^i$ is orthogonal to $d^i$ and $g^{ij}$ is independent from the angles,
\begin{equation}
\partial_a d^i = S^{ij} \partial_a d_j = \omega^{-1} S^{ij} \partial_a k_j,
\end{equation}
therefore
\begin{equation}
I^i_a \equiv \int^{\eta_\obs}_\eta \dd\eta' \; \partial_a d^i
=
\p{ \int^{\eta_\obs}_\eta \dd\eta' \; \omega^{-1} S^{ij}} \partial_a k_j
\equiv
\mathcal{I}^{ij} \partial_a k_j,
\end{equation}
where we used that $k_i$ is conserved along null geodesics.

Let us now start from the GLC expression~\eqref{eq:Jacobi_GLC} of the Jacobi matrix
\begin{equation}
\jacobi\indices{^A_B} = s^A_a \, C^a_B,
\end{equation}
where the inverse of the constant matrix reads $C^A_a = (\dd s^A_a/\dd\lambda)_\obs$. Reference~\cite{FPU} has proposed various techniques and explicit examples to determine the components $s_A^i$ of the Sachs basis with respect to the comoving basis~$\vect{\partial}_i$. There is no need to repeat them here. Once those components are found, the GLC ones are simply expressed as
\begin{equation}
s^A_a = \vect{g}(\vect{s}_A,\vect{\partial}_a)
= \vect{g}(\vect{s}_A,I^i_a \vect{\partial}_i) 
= s^A_i I^i_a,
\end{equation}
since $(\vect{\partial}_a)^i=\partial x^i/\partial\theta^a=I^i_a$. The components of the matrix $[C^A_a]$ then read
\begin{align}
\ddf{s^A_a}{\lambda} 
&= \left.\pd{}{\eta}\right|_{\eta_\obs,\theta,\ph} ( I^i_a s^A_i ) \\
&= (\partial_a d^i) s^A_i + I^i_a (\partial_\eta s^A_i).
\end{align}
On $\wl_\obs$, as $\eta=\eta_\obs$ the second term vanishes, so that
\begin{equation}
C^A_a = (s^A_i \partial_a k^i)_\obs.
\end{equation}
We now claim that the inverse~$C^a_A$ of the above matrix reads
\begin{equation}\label{eq:constant_matrix_Jacobi}
C^a_A = I^a_i \mathcal{I}^{ij} (s_{Aj})_\obs.
\end{equation}
Indeed, their contraction yields
\begin{align}
C^a_A C^A_b = I^a_i \mathcal{I}^{ij} \underbrace{(s_{Aj})_\obs (s^A_l)_\obs}_{(S_{jl})_\obs} (\partial_b k^l)_\obs;
\end{align}
besides, note that
\begin{equation}
\mathcal{I}^{ij} (S_{jl})_\obs
= \mathcal{I}^{ij} ( g_{jl} - d_j d_l )_\obs
=\mathcal{I}^{ij} (g_{jl})_\obs
\end{equation}
since $(d_i)_\obs=k_i$, and $\mathcal{I}^{ij}k_j=0$. Therefore, we have
\begin{align}
C^a_A C^A_b 
&= I^a_i \mathcal{I}^{ij} (g_{jl})_\obs (\partial_b k^l)_\obs \\
&= I^a_i \mathcal{I}^{ij} \partial_b k_j \\
&= I^a_i I^i_b \\
&= \delta^a_b.
\end{align}
Plugging the expression~\eqref{eq:constant_matrix_Jacobi} of $C^a_A$ in the GLC expression of the Jacobi matrix finally yields exactly eq.~\eqref{eq:Jacobi_FPU}, which concludes the proof.

\subsection{Angular diameter distance}

Following eq.~\eqref{eq:GLC_DA}, the angular distance in a Bianchi~I spacetime is given by
\begin{equation}
D\e{A}^2 
= \frac{a^2}{\sin\theta} 
	\sqrt{ \p{\sum_{i=1}^3 \ex{2\beta_i} (I^i_1)^2} \p{\sum_{i=1}^3 \ex{2\beta_i} (I^i_2)^2} 
				- \p{\sum_{i=1}^3 \ex{2\beta_i} I^i_1 I^i_2}^2 },
\end{equation}
where we used that $\gamma_{ab}=g_{ij} I^i_a I^j_b$. Though not so complicated, this expression requires to compute the six different integrals $(I^i_a)^{i=1,2,3}_{a=1,2}$, and is therefore less efficient than the alternative formula
\begin{equation}\label{eq:DA_Saunders}
D\e{A}^2 = a^2\,\tilde{\Upsilon}^{-1} \sum_{i\not=j\not=\ell} J_i J_j (k_\ell)^2
\end{equation}
with $k_\ell = e^{\beta_\ell^o} D^\ell$ from eq.~\eqref{eq:ki_tilde} and
\begin{equation}\label{eq:DefJi}
J_i \define \int_{\eta}^{\eta_\obs} \dd \eta' \; \tilde{\Upsilon}^3 \ex{2(\beta_i-\beta_i^\obs)},
\end{equation}
as found by Saunders in ref.~\cite{saunders_observations_1969} and recently rederived in refs.~\cite{2014MNRAS.444.2820S,FPU}, which involves only three different integrals ($J_i$). The equality between those two expressions for $D\e{A}$ is not to be proven, as our GLC expression of the Jacobi matrix agrees with the one of ref.~\cite{FPU}, which itself agrees with Saunders' formula~\eqref{eq:DA_Saunders}.

\subsection{Directional average of the inverse magnification}
\label{SecDirAverage}

A still ongoing debate in cosmology is whether or not gravitational lensing affects the average distance-redshift relation, that we use to interpret almost all cosmological observations~\cite{Fleury:2015hgz}. In particular, in ref.~\cite{1976ApJ...208L...1W} Weinberg argued that there is no such effect in a transparent Universe, a result physically justified by energy conservation. Although this claim and its justification are inaccurate---as already noticed by Bertotti~\cite{1966RSPSA.294..195B} and emphasized later on by several authors~\cite{Mustapha:1997xb,Ellis:1998ha,Ellis:1998qga,Enqvist:2009hn,Mustapha:1998jb,Enqvist:2007vb,GarciaBellido:2008gd,Bolejko:2011jc,Sundell:2015cza,Lavinto:2013exa}---a weaker version of it has been formulated by Kibble \& Lieu, who showed that in a locally inhomogeneous Universe the total area of a constant-affine-parameter surface is essentially unaffected by gravitational lensing~\cite{2005ApJ...632..718K}. This is equivalent to saying that the directional average of the \emph{squared} angular distance~$\ev{D\e{A}^2(\lambda)}_\Omega$ is unchanged with respect to its background value.\footnote{The area of an iso-$\lambda$ surface is indeed
\begin{equation}
A(\lambda) = \int_{\lambda} \dd^2 A = \int_{\lambda} D\e{A}^2(\lambda) \dd^2\Omega \define 4\pi\ev{D\e{A}^2(\lambda)}_\Omega,
\end{equation}
so that $A(\lambda)=A\e{FLRW}(\lambda)$ is equivalent to $\ev{D\e{A}^2(\lambda)}_\Omega=\pac{D\e{A}\h{FLRW}(\lambda)}^2$.
}
If moreover the affine parameter-redshift relation is \emph{assumed} to be essentially unaffected by the inhomogeneity of the Universe, then one obtains the conjecture
\begin{equation}
\ev{\mu^{-1}(z)}_\Omega
\define \ev{\pac{\frac{D\e{A}(z)}{D\e{A}\h{FLRW}(z)}}^2}_\Omega
= 1.
\label{eq:inverse_magnification_unity}
\end{equation}
This relation has been carefully checked at second-order in the cosmological perturbation theory, by the independent works of Bonvin et al.~\cite{2015JCAP...07..040B}, Kaiser \& Peacock~\cite{Kaiser:2015iia}, and Ben-Dayan et al.~\cite{P3,P5}. Note however that the non-perturbative numerical analysis of light propagation through a random Swiss-cheese model with transparent Lema\^itre-Tolman-Bondi holes, performed by Lavinto \& R\"as\"anen~\cite{2015JCAP...10..057L}, exhibited deviations with respect to this rule.

In this subsection, we show that the effect of anisotropy on cosmological  observations also violates the unity of $\ev{\mu^{-1}}_\Omega$. For simplicity, we consider from now on an axisymmetric Bianchi~I spacetime, with $(\beta_i)=(\beta,\beta,-2\beta)$, filled by dust and with no cosmological constant. We then compare the (squared) angular distance-redshift relation~$D\e{A}^2(z)$ in this anisotropic universe to the isotropic case. There are however different ways to draw a correspondence between a Bianchi~I and a FLRW model. Are considered here what we believe to be the two most sensible identifications: (i) models with the same volumic expansion rate~$H_0$ today; or (ii) models with the same matter density~$\rho_0$ today. In the former case $\rho_0\h{BI}/\rho_0\h{FLRW}=1-\Omega_{\sigma 0}$; in the latter case $H_0\h{FLRW}/H_0\h{BI}=\sqrt{1-\Omega_{\sigma 0}}$.

\begin{figure}[t]
\centering
\includegraphics[scale=1]{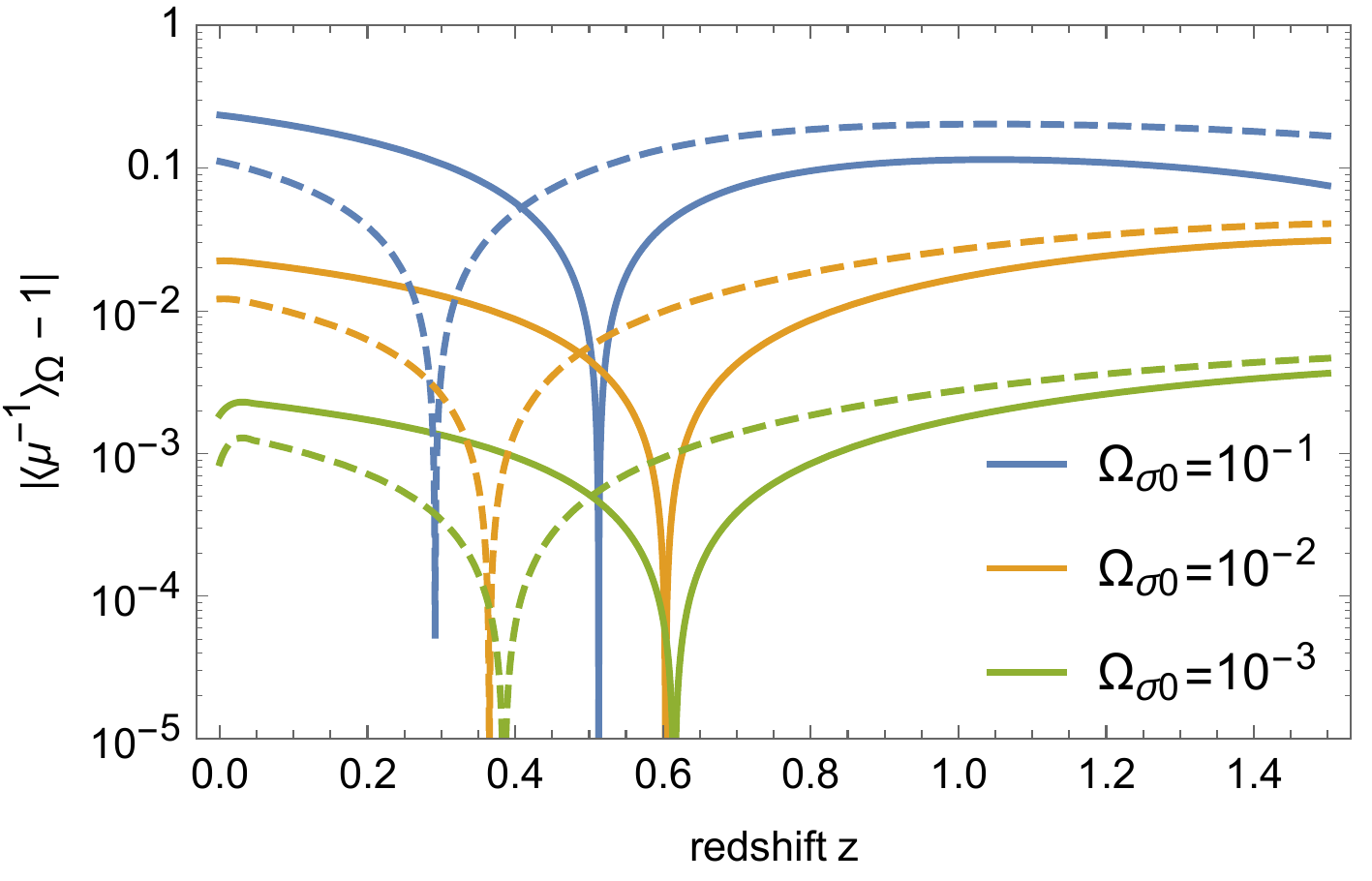}
\caption{Directional average of the inverse magnification in a Bianchi I universe with respect to a FLRW universe defined by $\mu^{-1}\define  (D\e{A}\h{BI}/D\e{A}\h{FLRW})^2$, as a function of redshift~$z$, for three different values of the anisotropy parameter~$\Omega_{\sigma 0}=10^{-1},10^{-2},10^{-3}$. There are two sensible ways to associate a FLRW model with a Bianchi~I model, either by identifying their Hubble constant today~$H_0$ (solid lines), or by identifying their matter density today~$\rho_0$ (dashed lines). For each curve $\ev{\mu^{-1}}_\Omega-1$ is first positive and then becomes negative.}
\label{fig:inverse_magnification}
\end{figure}

The results are plotted in fig.~\ref{fig:inverse_magnification}, for three different values of the anisotropy parameter~$\Omega_{\sigma 0}$. We see that, regardless of the identification scheme between anisotropic and isotropic models (same expansion rate or same density), we get $\ev{\mu^{-1}}_\Omega-1 \sim \Omega_{\sigma 0} \sim \sigma^2$. The violation of eq.~\eqref{eq:inverse_magnification_unity} is therefore \emph{of order two} in metric perturbations, while the analysis of ref.~\cite{2015JCAP...07..040B} suggests that $\ev{\mu^{-1}}_\Omega-1=\mathcal{O}(4)$. The fact that Bianchi I---which can be seen as a perturbation with respect to FLRW if $\Omega_{\sigma 0}\ll 1$---escapes from the eq.~\eqref{eq:inverse_magnification_unity} can be explained by two phenomena. First, the relation between affine parameter and redshift is much more affected in Bianchi~I than in a perturbed FLRW model, as the impact of anisotropy is cumulative, whereas most of the effects cancel in the statistically homogeneous and isotropic case. Second, an important identity leading to eq.~\eqref{eq:inverse_magnification_unity} in a perturbed FLRW universe is that the first-order convergence is on average equal to the first-order optical shear, $\ev{\kappa_1^2}=\ev{\gamma_1^2}$~\cite{2015JCAP...07..040B}. This has no reason to apply in Bianchi~I. Both effects combine in a way that eventually violates the conservation of the proper area of iso-$z$ surfaces.

These two differences between Bianchi~I and FLRW also explain the behaviour of $\ev{\mu^{-1}}_\Omega-1$ with redshift, namely positive [$\ev{(D\e{A}\h{BI})^2}_\Omega > (D\e{A}\h{FLRW})^2$] at low $z$ and then negative. At low redshift, anisotropic expansion manifests mostly as a modification of the Hubble law: the ratio between the recession velocity (hence redshift) and distance is not uniform in the sky but depends on the direction in which the source is observed. This effect of anisotropy is completely equivalent to the presence of a large-scale anisotropic peculiar velocity field within an otherwise isotropic Universe. It has been understood in ref.~\cite{2015MNRAS.454..280K} that any peculiar velocity field generates a positive (Malmquist-like) bias in the sky-averaged distance-redshift relation. The low-$z$ behaviour of $\ev{\mu^{-1}}_\Omega-1$ is thus in agreement with this general result. At larger redshifts, the effect of peculiar velocities decreases, progressively dominated by gravitational lensing. The fact that $\ev{[D\e{A}\h{BI}(z)]^2}_\Omega < [D\e{A}\h{FLRW}(z)]^2$ in this regime is due to the presence of Weyl lensing in Bianchi~I, which tends to focus light (see e.g. ref.~\cite{Fleury:2015hgz}), i.e. magnify images, and thus reduces the observed angular distance at a given redshift. This interpretation is also consistent with the fact that the turnover redshift (when $\ev{\mu^{-1}}_\Omega-1$ changes sign) is higher when the Bianchi~I and FLRW have the same $H_0$ (solid lines in fig.~\ref{fig:inverse_magnification}). Indeed, in this case the Bianchi~I model has a smaller matter density, $\rho_0\h{BI}=(1-\Omega_{\sigma 0}) \rho_0\h{FLRW}<\rho_0\h{FLRW}$ , so its Ricci focusing is weaker, which counterbalances the presence of Weyl lensing. Although the latter increases faster with redshift and eventually dominates, its overall effect is reduced and the turnover happens at higher $z$.

\subsection{Bias and dispersion of the Hubble diagram}

The violation of \eqref{eq:inverse_magnification_unity} presented in the previous subsection is mostly a conceptual issue. Let us now turn to a more observationally relevant impact of cosmic anisotropy, namely the bias and dispersion of the Hubble diagram of SNe, i.e. of the magnitude-redshift relation $m(z)$. Our goal is still essentially illustrative, hence we will not try to be realistic and will use the same setting as in subsec.~\ref{SecDirAverage}. The inclusion of a cosmological constant, as well as more realistic values for the anisotropy and its sources, shall be postponed to future works.

We remind the reader that the apparent magnitude~$m$ of a luminous source is defined from its observed luminosity in such a way that
\begin{equation}
m = 5\log_{10} D\e{L} + \cst ,
\end{equation}
where $D\e{L}$ is the luminosity distance between the source and the observer, related to $D\e{A}$ by the distance-duality relation~\cite{Fleury:2015hgz} $D\e{L}=(1+z)^2 D\e{A}$, and where the constant only depends on the intrinsic properties of the source. The Hubble diagram consists in plotting the magnitude (more precisely the distance modulus) of a catalogue of SNe as a function of their redshift. In a homogeneous and isotropic FLRW universe, this relation is univocal, in particular it does not depend on the position of the SN on the sky. In a Bianchi~I universe, however, it does depend on the direction so that the Hubble diagram gets both biased and spread with respect to the isotropic case.

Because the Hubble diagram is constructed from individual sources, the relevant statistics involves averages over sources~$\ev{\ldots}_N$ defined in subsec.~\ref{subsec:LCaverage}, rather than directional averages. The bias of the Hubble diagram is therefore
\begin{equation}
\Delta m(z) \define \ev{m\e{BI}(z)}_N - m\e{FLRW}(z),
\end{equation}
while its dispersion can be quantified by
\begin{equation}
\sigma_m(z) \define \sqrt{ \ev{m\e{BI}^2(z)}_N - \ev{m\e{BI}(z)}_N^2 }.
\end{equation}
Given the Bianchi~I model considered here, it is straightforward to show that the source average defined by eq.~\eqref{eq:SourceAverage} takes the form
\begin{equation}
\ev{S}_N(z) = 
\frac{\int \dd^2\Omega \, K_N(z,\theta) \, S}{\int \dd^2\Omega \; K_N(z,\theta) },
\end{equation}
with the integration kernel
\begin{equation}
K_N(z,\theta) =  D\e{A}^2(z,\theta) 
\pac{ \frac{a_\obs}{a(z)} }^3
\paac{ H(z) + \frac{\sum_{i=1}^3 \dot{\beta}_i(z) \pac{\ex{\beta_i^\obs-\beta_i(z)} D^i(\theta,\ph)}^2}
								{\sum_{i=1}^3 \pac{\ex{\beta_i^\obs-\beta_i(z)} D^i(\theta,\ph)}^2}
		}^{-1}.
\end{equation}
Note that $K_N$ does not depend on $\ph$ because of the axisymmetry of our Bianchi~I spacetime.

\begin{figure}[t]
\centering
\includegraphics[width=0.49\columnwidth]{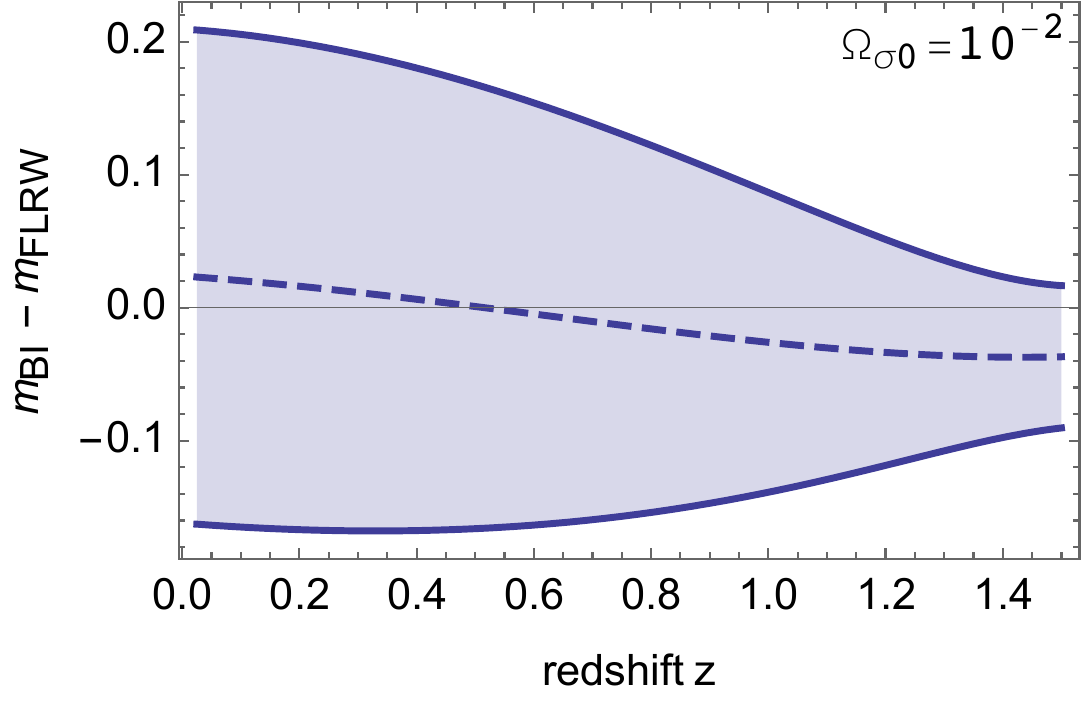}
\hfill
\includegraphics[width=0.49\columnwidth]{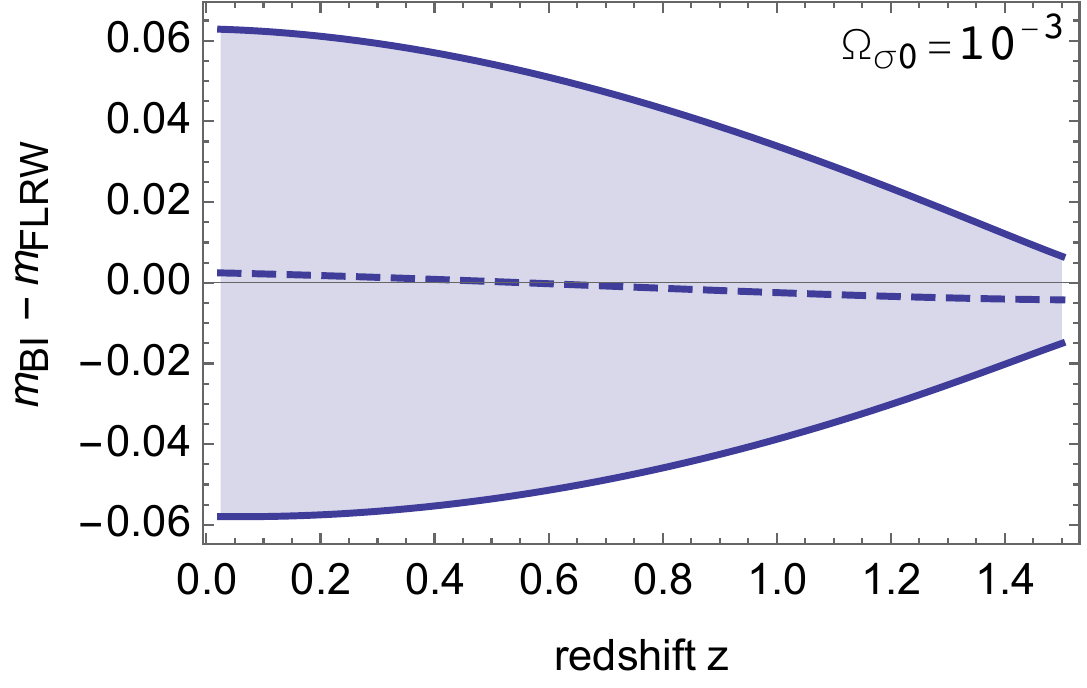}
\caption{Bias and dispersion of the Hubble diagram of SNe due to anisotropy, with respect to an isotropic model with the same volumic expansion rate $H_0$ today. Dashed lines represent the bias~$\Delta m\define \ev{m\e{BI}}_N-m\e{FLRW}$, while solid lines represent $\Delta m \pm \sigma_m$. Two values of the anisotropy parameter are considered: $\Omega_{\sigma 0}=10^{-2}$ (left panel) and $\Omega_{\sigma 0}=10^{-3}$ (right panel).}
\label{fig:HubbleDiagram}
\end{figure}

The behaviours of $\Delta m$ and $\sigma_m$ are depicted in fig.~\ref{fig:HubbleDiagram}, for two different values of the anisotropy parameter, namely $\Omega_{\sigma 0}=10^{-2},10^{-3}$. In terms of orders of magnitude, the bias is $\Delta m\sim\Omega_{\sigma 0}$, while the dispersion is $\sigma_m\sim10\,\Omega_{\sigma 0}$. For comparison, the typical dispersion of the most recent Hubble diagram~\cite{2014A&A...568A..22B} is of order $0.2\U{mag}$. Note the interesting shrinking of this dispersion at high redshift, due to the fact that the distance-redshift relations~$D\e{A}(z,\theta)$ in various directions~$\theta$ tend to approximately intersect. Such a behaviour is a feature of anisotropy which cannot be mixed with any other effect, such as lensing or peculiar velocities~\cite{P3}. This could be exploited in a component-separation analysis in order to efficiently detect an anisotropy signal from the Hubble diagram with future large SN surveys.

It may seem far fetched to try to constrain cosmic anisotropy from the Hubble diagram, while extremely tight constraints have been put by the lastest analyses of the CMB~\cite{2016arXiv160401024S,2016arXiv160507178S}, namely $\Omega_{\sigma 0}<10^{-13}$ (and even $<10^{-20}$ for scalar and vector anisotropy modes). However, such constraints are based on the analysis of the polarisation of the CMB, and thus probe \emph{primordial} anisotropy, which decays later on if not sourced. This method is much less sentitive to \emph{late-time} anisotropy, which is a prediction of many phenomenological dark energy model~\cite{Bucher:1998mh,Mota1,Mota3,Appleby2,Appleby3}, bi-gravity theories~\cite{bigrav}, or backreaction~\cite{marozzi}. Using low-redshift observations such as SNe is therefore complementary to the analyses reported in refs.~\cite{2016arXiv160401024S,2016arXiv160507178S}.

\section{Conclusion}
\label{sec:conclusion}

The geodesic-light-cone (GLC) coordinate system has been introduced and exploited in several previous works in order to facilitate the analysis of cosmological observations in an inhomogeneous universe~\cite{P1,P2,P3,P4,P5,P6,Marozzi:2014kua,Fanizza:2015swa,DiDio:2014lka}. In this article, we presented the first bottom-up construction of those coordinates, explicitly derived the expression of the GLC metric, and explored its gauge degrees of freedom in details. We also reminded the reader the form of the equations of geometric optics in terms of the GLC coordinates, and clarified the general expression of the Jacobi matrix by providing a simple interpretation of the constant matrix~$[C^a_A]$.

The second part of the article was dedicated to the application of GLC coordinates to the homogeneous and anisotropic Bianchi~I cosmological model. We derived the expression of the associated metric, and checked the consistency of its optical predictions with the literature. We finally exploited our results to analyse the statistical properties of the distance-redshift relation of an anisotropic universe. On the one hand, we showed that the directional average of the inverse magnification with respect to a FLRW model \emph{differs from 1}, $\ev{\mu^{-1}(z)}_\Omega\not= 1$, in other words, the Weinberg conjecture stating that the area of an iso-$z$ surface is unaffected by gravitational lensing is violated in a Bianchi~I spacetime. In that respect, the effect of anisotropy is thus different from the one of inhomogeneity, as in the latter case the inverse magnification is shown to be unity when averaged over the sky~\cite{2015JCAP...07..040B,Kaiser:2015iia,P3,P5}. On the other hand, we illustrated the effect of anisotropy on the Hubble diagram, which displays a bias on the order of $\Omega_{\sigma 0}\U{mag}$ and a dispersion on the order of $10\,\Omega_{\sigma 0}\U{mag}$. The particular behaviour of this dispersion as a function of the redshift, quite distinct from the effects of inhomogeneity, arises as an interesting property to extract any anisotropy signal from future SN observations.

This last point clearly calls for more observationally-oriented follow-up works. It could be complemented by other lensing observables, such as the cosmic shear generated by anisotropic expansion. Furthermore, the theoretical results presented here should facilitate the analysis of light propagation through anisotropic \emph{and} inhomogeneous cosmological models---such as perturbed Bianchi~I spacetimes---in the same spirit of what has been already done with perturbed FLRW spacetimes. Finally, the whole analysis presented here could be repeated for the other types of Bianchi models.


\section*{Acknowledgements}
PF and GF thank Giovanni Marozzi for stimulating debates on the GLC coordinates during the early stages of this project; PF thanks Chris Clarkson for clarifying discussions on averages in gravitational lensing; and we all warmy thank Chris Clarkson, Roy Maartens, Giovanni Marozzi, Syksy R\"as\"anen, and Gabriele Veneziano for valuable comments on a preliminary version of this article. PF's research is funded by the South African National Research Foundation. The financial assistance of the National Research Foundation (NRF) towards this research is hereby acknowledged. Opinions expressed and conclusions arrived at, are those of the author and are not necessarily to be attributed to the NRF. FN's research was supported by the project GLENCO (FP7, Ideas, Grant Agreement n. 259349) during the first stages of this project. His research is now supported by the Leung Center for Cosmology and Particle Astrophysics (LeCosPA) of National Taiwan University. GF's work is supported by the research grant ``Theoretical Astroparticle Physics'' number 2012CPPYP7 under the program PRIN 2012 funded by the Ministero dell'Istruzione, Universit\`{a} e della Ricerca (MIUR) and by the Italian Istituto Nazionale di Fisica Nucleare (INFN) through the ``Theoretical Astroparticle Physics'' project. GF also wish to thank the University of Geneva for its hospitality during part of the elaboration of this document.

\bibliographystyle{JHEP.bst}
\bibliography{bibliography}


\end{document}